\documentclass[a4paper,11pt]{article}
\usepackage[utf8]{inputenc}
\usepackage{xspace, times, graphicx, cite, url}
\usepackage{geometry}
\usepackage{cogabbr, textshortcuts}
\usepackage{caption}
\title{\textbf{Optical coherence tomography --- principles, implementation, and applications in ophthalmology}}
\author{\textbf{Yoshiaki Yasuno}\\
	\textit{Computational Optics Group, University of Tsukuba, Tsukuba, Japan.}\\
	yoshiaki.yasuno@cog-labs.org\\
	https://optics.bk.tsukuba.ac.jp/COG/\\
	https://cog-news.blogspot.com/}
\date{Ver.\@ 3.0a, Dec. 8, 2022.}

\graphicspath{{./}{./figures/}}	
 
\begin{document}
 \maketitle
 
\begin{abstract}
This short textbook was written for someone who newly start studying, doing research, or development about optical coherence tomography (OCT) or using OCT. 
The first chapter summarizes the concept and the history of OCT.
In the second chapter, the technologies of OCT are summarized, which includes the mathematical principle, hardware implementation, the instructions for designs such as interferometer and wavelength selections, and theories of resolutions and sensitivity.
In the third chapter, some examples of ophthalmic applications of OCT are described.
In the last chapter, instructions for next learning steps are given.
\end{abstract}

\tableofcontents

\section{Introduction}
\subsection{What is optical coherence tomography?}
Optical coherence tomography (OCT) is an interferometric modality that
provides noninvasive tomography of \invivo human tissues \cite{Huang1991Science}.
OCT is essentially a scanning low-coherence interferometer that utilizes coherence
gating to resolve the depth structure of a sample.
During the measurement, OCT illuminates a sample with a probe beam focused on it.
The transversal structure is then obtained by transversally scanning the probe beam using a rotating mirror, typically a galvanometric scanning mirror.
If the transversal scanning is one-dimensional (1-D), a 2-D cross-sectional image is obtained.
Similarly, 2-D transversal scanning, typically achieved by a pair of galvanometric scanning mirrors, provides a 3-D volumetric tomography.

Among several medical tomographic modalities, OCT is characterized by the following properties. First, OCT is noninvasive.
Since OCT uses a weak near-infrared (NIR) light as a probe, the sample remains free of
photochemical and photothermal damage.
Second, the OCT has higher resolution than other clinical tomographic methods such as X-ray computed tomography, magnetic resonance imaging, and ultrasound tomography \cite{IntroOCT}.
OCT typically provides a depth resolution of less than 10 \um.
Third, the measurement speed of OCT is fast. The first generation of OCT, known as
time-domain OCT (TD-OCT), had an imaging speed between several hundreds to several thousands of depth lines (A-lines) per second  \cite{Rollins1998OpEx}.
A more recent OCT method, Fourier-domain OCT (FD-OCT) provides an even faster
speed of around several hundred-thousand A-lines per second \cite{Potsaid2010OpEx, Klein2013BOE}.
This high speed enables 3-D investigation of \invivo organs within a realistic
measurement time.
Another characteristic property of OCT is its maximum measurement depth or penetration depth.
Since OCT uses NIR light, the probe beam is relatively highly scattered by the sample to be measured.
This high scattering unfortunately limits the penetration depth of the OCT to
within a few millimeters.
Considering these properties of OCT, it is clear that OCT is not an all-purpose tomographic modality.
However, there are some organs that are extremely suitable for OCT.

\subsection{OCT and ophthalmology}
\subsubsection{The epoch of OCT}
OCT has been utilized for variety of clinical fields such as dermatology \cite{Pierce2004JID}, dentistry \cite{Otis2000JADA}, gastrointestinal medicine \cite{Gora2013NatMed}, oncology \cite{Vakoc2012NRC}, cardiology \cite{Tearney2008JACCI}, and ophthalmology \cite{Sakata2009CXO}.
Among these clinical fields, ophthalmology is perfectly suited to OCT.

The first demonstration of OCT was described in a research paper
published in 1991 \cite{Huang1991Science}.
In this seminal paper, it is clear that OCT was intended to be utilized in ophthalmology from the beginning.
This intention can be inferred from three facts. First, the samples measured in this paper were a macular and an optic nervehead of an \exvivo porcine retina.
Structural disorder of the macula is associated with several severe eye diseases such as age-related macular degeneration (AMD), diabetic retinopathy, and central
serous chorioretinopathy (CSC), while the abnormality of the optic nervehead
is associated with glaucoma.
This fact suggests that the authors, i.e., the inventors of OCT, already intended to use OCT for clinical practice in ophthalmology even at its epoch.

The second fact implicitly suggesting the authors' intention is the
authorship of the paper that comprises laser specialists, including James G.\@
Fujimoto from the Massachusetts Institute of Technology, and ophthalmology
specialists, including Joel S.\@ Schuman and Carmen A.\@ Puliafito from the Department of Ophthalmology, Harvard Medical School.
In addition, the first author, David Huang, was a PhD student studying both engineering and medicine.

The third fact concerns the series of papers that were quickly published by the authors after this first publication.
In 1992, the same authors published a high-speed low-coherence reflectometry method \cite{Swanson1992OL}.
It should be noted that a high imaging speed is crucial for \invivo OCT imaging
of an eye.
The submission date of this paper was September 1991, suggesting that the authors were preparing for \invivo OCT eye imaging even when they were working on their first 1991 publication.
Furthermore, in 1993, the authors demonstrated the first \invivo human retinal imaging \cite{Swanson1993OL}.

These facts suggest that OCT was, from the beginning, developed for use in clinical ophthalmology.
As proof of this speculation, OCT was commercialized as a clinical ophthalmic imaging device only 5 years after the first demonstration.

Regarding this point, the question arises as to why they were interested in ophthalmology.

\subsubsection{Eye is a perfect target of OCT}
The eye is a perfect organ for investigation by optical modalities.
This is mainly because the eye itself is an optical instrument.
Because the eye is an imaging system that images an object on the retina, it is also easy to build an optical system that images the retina on an arbitrary imaging plane using the eye optics.

Another important characteristic of the eye is transparency.
As discussed above, OCT penetration is limited to a few millimeters, mainly because of scattering in tissue.
However, we can ignore the scattering in the optical media of the eye including the cornea, aqueous humor (that fills the anterior eye chamber), crystalline lens, and vitreous.
Hence, several-millimeter penetration of retinal OCT is defined not from the surface of the eye but from the surface of the retina, located around 24-mm under the eye surface.
Because the thickness of the retina is about a few hundred micrometers, the limited penetration of OCT is not a serious issue for retinal imaging.

The direction of the tomographic cross section also enhances the value of OCT.
Almost all eye imaging modalities such as ophthalmoscope, color fundus photography, and angiographies are en-face imaging modalities, while OCT is the only one that is an \invivo modality that provides a depth-oriented cross section of the retina.
Because the retina has a fine multi-layered structure and its destruction is highly associated with eye diseases, the \invivo cross-sectional tomography provided by OCT is extremely important for ophthalmic diagnosis.

The non-invasiveness of OCT is also particularly important in ophthalmology.
This is because the eye is one of a few organs that we cannot biopsy.
Since all portions of the retina are associated with visual function, it is impossible to excise even a small portion of a tissue for diagnosis purposes.

Considering these properties of the eye, it is natural to conclude that the
eye is a perfect organ for investigation by OCT.
Furthermore, it is natural that the eye was selected as the first target for OCT.

\section{OCT technologies}
\subsection{Requirement for eye imaging}
The first generation of OCT, TD-OCT, is a variant of low-coherence interferometry.
Hence, TD-OCT is a relatively slow imaging modality.
TD-OCT scans the optical path-length difference for depth sectioning and scans
the probe beam transversally to obtain transversal structural information.
In particular, a 2-D scan is required for 2-D cross-sectional imaging, and a 3-D
scan is required for volumetric tomography.
This multi-dimensional mechanical scanning limits the measurement speed.
For example, one widely utilized retinal OCT device (Stratus OCT, Carl Zeiss Meditec, CA, USA) requires 1.5 s to obtain a single cross-sectional scan (B-scan).
The A-line (depth scan) rate of this device is only approximately 340 A-lines/s.

Measurement speed is particularly important for \invivo eye imaging because the eye is always moving.
Involuntary eye motions including drift, tremor, saccade, and micro-saccade are normal physiological actions, and hence inevitable.
The OCT measurement speed needs to be fast enough to render eye motion negligible.
The above-mentioned speed of 340 A-lines/s was somewhat acceptable for 2-D cross-sectional imaging but not fast enough for volumetric measurement.
In fact, the retinal cross-section taken by clinical TD-OCT was significantly distorted by eye motion.
Namely, TD-OCT was not sufficiently fast enough, even for 2-D imaging.
A state-of-the-art TD-OCT has demonstrated a higher A-line rate of 4,000 A-lines/s \cite{Rollins1998OpEx}, but this speed was still not sufficiently fast for volumetric imaging.
In order to obtain a volumetric tomography of the eye, a measurement speed on the order of tens of thousands of A-lines/s is required.

Another important requirement for ophthalmic OCT is high sensitivity.
Because the reflectivity of retina can be as low as $10^{-5}$ to $10^{-6}$, a high system sensitivity and the resulting high signal-to-noise ratio (SNR) are
required.
In general, the SNR of OCT is roughly proportional to the measurement time and optical power of the probe beam.
As discussed above, the OCT measurement time should be short for eye investigation.
Hence, it is not a good strategy to use long measurement times to increase the SNR.
Increasing the probe beam power is also not an optimal solution.
Because the eye is a photosensitive organ, it is also very sensitive to photodamage. Hence, the probe beam power of retinal OCT is strictly limited by safety standards
such as ANSI \cite{ANSI136.1-2014} and ISO \cite{ISO15004-1-2009}.
For example, the maximum allowable probe power for OCT is around 700 \uW for a probe wavelength of 840 nm and around 1.2 to 1.7 mW for 1060 nm.
To overcome these inherent sensitivity limitations, a new OCT method that had a higher probe beam throughput for image formation was required.

These two main requirements, i.e., high-speed and high-sensitivity, have driven researchers to develop a new OCT variation: FD-OCT.

\subsection{TD-OCT versus FD-OCT}
\subsubsection{Advantages of FD-OCT over TD-OCT}
Although FD-OCT is a variant of OCT, it is based on a completely different interferometric configuration.
Namely, FD-OCT detects the interference signal in the spectral domain.
This spectral domain detection is typically achieved by a high-speed spectrometer or by a high-speed wavelength
sweeping laser that is equipped with an intracavity monochrometer.
FD-OCT based on the former scheme is called spectral domain OCT (SD-OCT) or spectral radar \cite{Fercher1995OC, Hauesler1998JBO}, and that based on the latter scheme is swept-source OCT (SS-OCT) or optical frequency domain imaging \cite{Yun2003OFDI}.

\begin{figure}[tbh]
	\centering\includegraphics{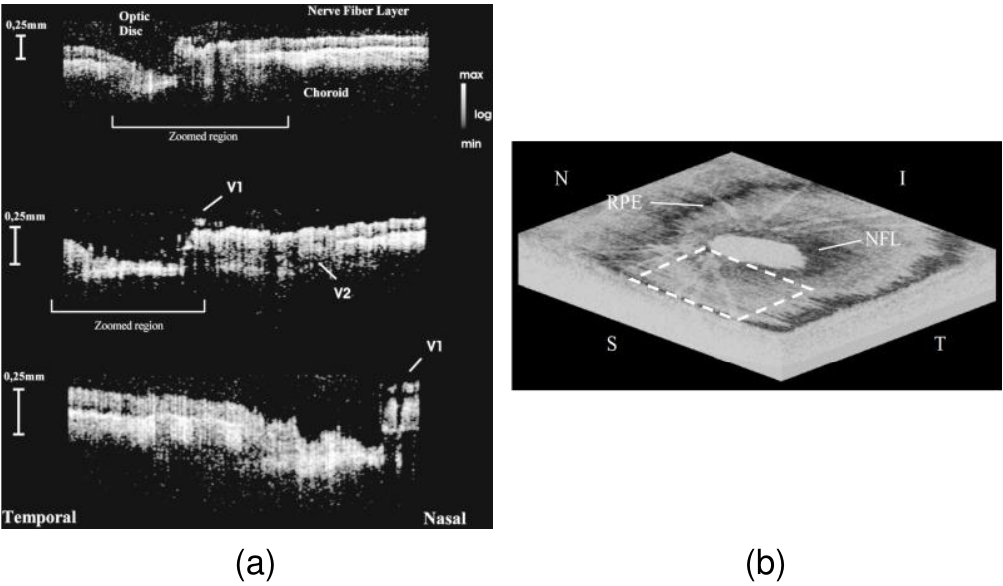}
	\caption{(a) The first \invivo SD-OCT image of a human retina.
		V1 and V2 indicate blood vessels.
		(b) Example of an \enface resection of a 3-D volume of a human retina obtained by a high-speed SD-OCT.
		The labels I, S, N, and T, respectively represent the inferior, superior, nasal, and temporal directions.
		RPE and NFL indicate the retinal pigment epithelium and nerve fiber layer, respectively.
		The images are reprinted from Refs.\@ \citen{Wojtkowski2002JBO} (a) and \citen{Nassif2004OpEx} (b).
	}
	\label{fig:firstRetinalOct}
\end{figure}
Because SD-OCT provides an A-scan without mechanical scanning, it achieves a faster measurement speed than TD-OCT.
The first 2-D \invivo retinal imaging using SD-OCT was performed with a measurement speed of 1,000 A-lines/s in 2002 \cite{Wojtkowski2002JBO} (Fig.\@ \ref{fig:firstRetinalOct}(a), reprinted from Ref.\@ \citen{Wojtkowski2002JBO}) and the first 3-D \invivo retinal image was achieved in 2004 with a measurement speed of 29,000 A-lines/s \cite{Nassif2004OpEx} (Fig.\@ \ref{fig:firstRetinalOct}(b), reprinted from Ref.\@ \citen{Nassif2004OpEx}).
Although SS-OCT requires an intracavity mechanical wavelength scanning mechanism such as
a piezoelectric Fabry–Perot filter, a polygon-mirror-based wavelength tuner, or a MEMS wavelength tuner, it provides a measurement speed of several ten- to hundred-thousand A-lines/s \cite{Yun2003OFDI}.

FD-OCT first attracted researchers because of its fast measurement speed.
However, soon it was recognized that it had a higher sensitivity than TD-OCT \cite{Leightgeb2003OpEx, deBore2003OL, Choma2003OpEx}.
This higher sensitivity can be seen as a result of the higher throughput of the probe beam to the image.
TD-OCT is a low-coherence interferometer in which the reference and probe beams generate an interference signal only when the optical-path-length difference (OPLD) is less than the coherence length.
Although this is the cause of the depth resolution of TD-OCT, it also means that the probe beam that is outside of the coherence length cannot contribute to image formation. On the other hand, FD-OCT is considered to be a bundle of monochromatic interferometers.
Namely, in the case of SD-OCT, the signal output at each wavelength channel of the spectrometer provides monochromatic interference between the reference and probe beams. Hence, the beams generate an interference signal even when the OPLD is several millimeters long. The broadband interference, i.e., low-coherence interference, which provides the depth resolution, is then numerically performed later in a computer. Because of this interference scheme, almost all portions of the probe beam power contribute to image formation.
The detailed theory that supports this high sensitivity is described elsewhere \cite{Leightgeb2003OpEx, deBore2003OL, Choma2003OpEx}.

Because of these two major advantages of FD-OCT, i.e., high speed and high sensitivity, it has quickly become the de facto standard for
tomographic investigation of the eye.

\subsection{Principles of FD-OCT}
\label{sec:principle}
\subsubsection{Hardware scheme of SD-OCT}
\begin{figure}[tbh]
	\centering\includegraphics{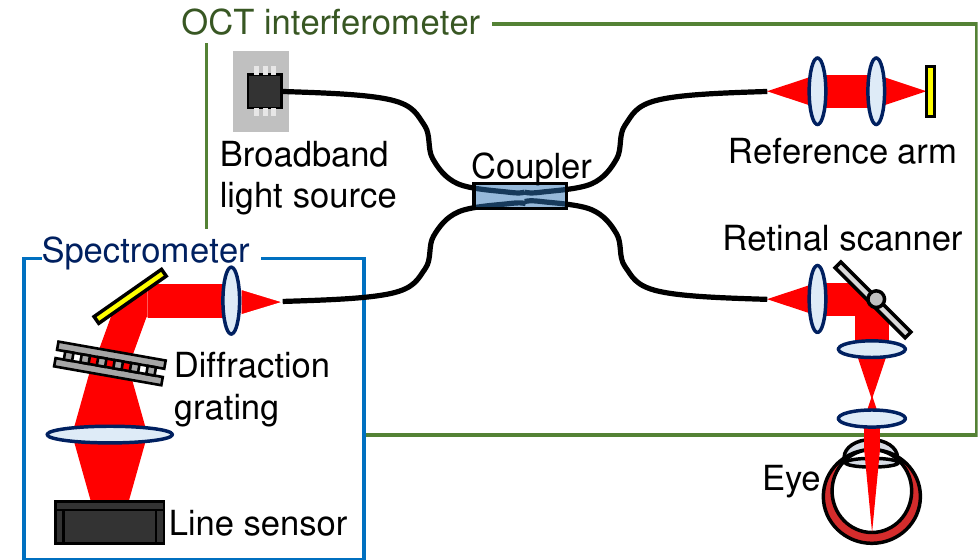}
	\caption{Simplified SD-OCT scheme consisting of two sub-systems, i.e., the OCT interferometer that is almost equivalent to that of TD-OCT and a high-speed spectrometer.
	The light source is a broadband, i.e., temporally incoherent light source.
	The beam from the light source is split by a fiber coupler into probe and reference beams.
	The probe beam illuminates the sample through a retinal scanner, which includes a pair of galvanometric scanning mirrors and an objective lens.
	The probe and reference beams are back-scattered by the sample and reflected by the reference mirror, respectively, combined by the coupler, and then introduced into the high-speed spectrometer.
	These two beams are then decomposed into its spectral components and form an interference signal in the wavelength spectral domain.}
	\label{fig:simpleScheme}
\end{figure}
In this section, SD-OCT is described as a representative FD-OCT method.
An SD-OCT system comprises two subsystems, as depicted in Fig.\@ \ref{fig:simpleScheme}.
The first subsystem is the OCT interferometer including the retinal scanner.
The second subsystem is a high speed spectrometer that typically consists of collimation optics, an optical grating, a focusing lens (Fourier transform lens), and a high-speed line sensor.
The line sensor is typically a 1-D CCD or CMOS camera.

The light source in the interferometer is a broadband light source that can be a broadband pulse laser such as Ti:Sapphire laser or a superluminescent diode (SLD).
A typical center wavelength is around 830 nm, and the bandwidth is from about 50 nm to more than 100 nm.
The reasons for this typical center wavelength and bandwidth are described later in Section \ref{sec:wavelength}.
The light from the light source is split by a fiber coupler.
A portion of the light is introduced into a probe arm that is equipped with a retinal scanner.
This portion is called the probe beam.
The residual portion of the beam is introduced into the reference arm of the interferometer and becomes the reference beam.
The backscattered probe beam and reflected reference beam are recombined by the coupler and then introduced into the high-speed spectrometer.
These two beams interfere with each other and create an interferogram, not in the time domain but in the spectral domain.
The spectral interference signal is then detected by a line sensor and numerically processed in a PC to form an OCT A-line.
A pair of galvanometric scanning mirrors in the retinal scanner is additionally utilized to obtain a 2-D or 3-D OCT tomography.

\subsubsection{Tomography reconstruction}
The numerical process to reconstruct an OCT A-line from the spectral interferogram consists of three phases: rescaling, inverse-Fourier transform, and logarithmic compression.

The first phase of the numerical process is rescaling.
It mainly rescales the spectral interferogram into the wavenumber domain.
The raw spectral interferogram detected by the high-speed spectrometer is almost linearly spaced in wavelength.
In the rescaling process, the spectral interferogram is numerically resampled to be linear in wavenumber. 

The rescaling process is further divided into two steps: calibration and resampling.
In the calibration step, the wavelength spacing property of the spectrometer is characterized.
The required accuracy of the wavelength calibration is significantly higher than that of standard spectrometers and can be as high as 0.1 nm or more.
Thus, the standard calibration methods of a spectrometer cannot be utilized.
On the other hand, some fringe analysis based methods can be used \cite{Yasuno2005OpEx, Makita2008OpEx}.
This calibration is required only once per spectrometer.
In the resampling process, the measured spectral interferogram is numerically resampled at a set of new sampling points that are evenly spaced in wavenumber.
In this numerical resampling, the spectral interferogram is interpolated by a polynomial/spline function.
It is also common to apply zero-padding interpolation before the polynomial/spline interpolation \cite{Nassif2004OpEx}.
The purpose of the zero-padding is to extend the inherent cut-off frequency of the polynomial/spline interpolation \cite{Dorrer2000JOSAB}.
The resampling is applied to all spectral interferograms obtained by the spectrometer.

The second phase of OCT A-line reconstruction is the inverse-Fourier transform.
This part is mathematically described as follows.
The electric field output from the OCT interferometer, and similarly that on the line sensor in the spectrometer, is described in the wavenumber domain as
\begin{equation}
	\label{eq:edk}
	\tilde{E_d}(k) = \sqrt{p_p \rho_p}\tilde{s}(k)\tilde{\alpha}(k)\tilde{\gamma_p}(k)
	+ \sqrt{p_r \rho_r}\tilde{s}(k)\tilde{\gamma_r}(k),
\end{equation}
where $k$ is the angular wavenumber, $p_p$ and $p_r$ are the optical powers of the probe and reference beams, respectively, $\rho_p$ and $\rho_r$ are the intensity throughput of the probe and reference arms, respectively, and $\tilde{s}(k)$ is the complex spectrum of the light source.
$\tilde{\gamma_p}$ and $\tilde{\gamma_r}$ are phase-only functions representing the group delay and the dispersion at the probe and sample arms, respectively.
The complex spectral reflectivity of the sample is $\tilde{\alpha}(k)$, and is defined using the complex reflectivity of the sample along depth $\alpha(z)$ as $\tilde{\alpha}(k) \equiv\ft{\alpha(z)}$ where $\ft{\quad}$ represents Fourier transform and $z$ is a variable representing the round trip optical path length (and is the Fourier counterpart of  $k$).
By assuming a refractive index of $n$, the depth position in the probe and reference arms $z'$ is expressed as $z'=z/2n$.
Note that because $\alpha(z)$ can be regarded as the structure of the sample along the depth, this is what we wish to measure.
Thus, the following processes aim to retrieve $\alpha(z)$ from $\tilde{E_d}(k).$

Although the spectrometer can perform spectrally resolved detection, the electric field $\tilde{E_d}(k)$ cannot be measured directly.
This is because the line sensor is not sensitive to the electric field but to light intensity.
Hence, we measure the power spectrum as
\begin{equation}
	\label{eq:idk}
	\begin{array}{rcl}
		\tilde{I_d}(k) &=& \left|\tilde{E_d}(k)\right|^2  \\
		&=& p_p'\left|\tilde{s}(k)\right|^2 \left|\tilde{\alpha}(k)\right|^2
		+ p_r' \left|\tilde{s}(k)\right|^2\\
		&&  +\sqrt{p'_p p'_r}\left|\tilde{s}(k)\right|^2\tilde{\alpha}(k)
		\tilde{\gamma_p}(k) \tilde{\gamma_r\conj}(k) + c.c.
	\end{array},
\end{equation}
where $p'_p \equiv p_p \rho_p$ and $p'_r \equiv p_r \rho_r$, $\tilde{I_d}(k)$ is the power spectrum, the superscript $*$ represents the complex conjugate, and $c.c.$ represents the complex conjugate of the third term.
Here, we have utilized $\left|\tilde{\gamma_p}(k)\right|^2 = \left|\tilde{\gamma_r}(k)\right|^2 = 1$ , i.e., $\tilde{\gamma_p}(k)$ and $\tilde{\gamma_r}(k)$ are phase-only functions.
It should be noted that the raw spectrum spread on the line sensor is not evenly spaced in wavenumber, but it is rescaled to the wavenumber domain by the previously described rescaling process.

\begin{figure}
	\centering\includegraphics{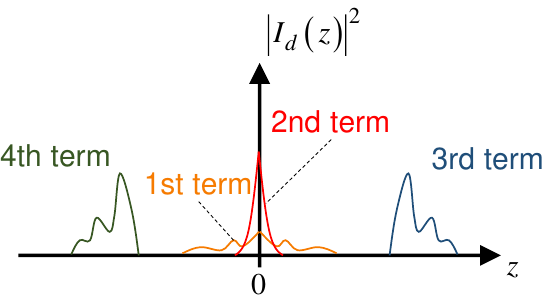}
	\caption{Schematic of the Fourier-transformed power spectrum.
	The first and second terms of Eq.\@ (\ref{eq:Idz2}) are appeared at the center, and correspond to the autocorrelation of the probe beam (first term) and the autocorrelation of the reference beam (second term).
	The third term is appeared at the right, and it is the cross-correlation between the probe and reference beams.
	This cross-correlation is the OCT signal. 
	The fourth term is a mirrored complex conjugate of the third term, and appeared at the left.
	This signal is frequently denoted as a mirror signal or a coherence artifact.
	}
	\label{fig:schemOctSig}
\end{figure}
The measured spectrum is then inverse-Fourier transformed as
\begin{equation}
	\label{eq:idz}
	\begin{array}{rcl}
		I_d(z)&\equiv &\ift{\tilde{I_d}(k)} \\
		&=&p'_p\Gamma(z) * \ac{\alpha(z)} + p'_r\Gamma(z)\\
		&&+ \sqrt{p'_p p'_r}\Gamma(z)*\alpha(z)*
		\ift{\tilde{\gamma_p}(k)\tilde{\gamma_r}\conj (k)}\\
		&&+ \sqrt{p'_p p'_r}\Gamma(z)*\alpha\conj(-z)*
		\ift{\tilde{\gamma_p}\conj(k)\tilde{\gamma_r} (k)}
	\end{array},
\end{equation}
where $\ift{\quad}$ represents inverse-Fourier transform, the operator $*$ represents convolution, $\Gamma(z)$ is the coherence function of the light source, defined as $\ift{\left|\tilde{s}(k)\right|^2}$, and $\ac{\alpha(z)}$ is the auto-correlation of $\alpha(z)$ which is, according to the Wiener-Khinchin's theorem, equivalent to $\ift{\left|\tilde{\alpha}(k)\right|^2}$.
Assuming the probe and reference arms have the same high-order phase dispersion, $\phi(k)$, the phase-only functions become $\tilde{\gamma_p}(k)=\sexp{i\phi(k)}\sexp{-ikz_p}$ and $\tilde{\gamma_r}(k)=\sexp{i\phi(k)}\sexp{-ikz_r}$, where $z_p$ and $z_r$ are the round-trip path length of the probe and reference arms, respectively. By substituting these phase terms into Eq. (\ref{eq:idz}), the Fourier-transformed power spectrum becomes
\begin{equation}
	\begin{array}{rcl}
		\label{eq:Idz2}
		I_d(z)&= &p'_p \Gamma(z) * \ac{\alpha(z)}+ p'_r\Gamma(z)\\
		&& + \sqrt{p'_p p'_r}\Gamma(z)*\alpha(z)*\delta(z-z_d)
		+ \sqrt{p'_p p'_r}\Gamma(z)*\alpha\conj(-z)*\delta(z+z_d)
	\end{array}
\end{equation}
where $z_d \equiv z_p - z_r$.
This signal is schematically depicted in Fig.\@ \ref{fig:schemOctSig}.
In this figure, the horizontal axis represents the double pass depth position $z$ and the vertical axis represents the squared power of  $I_d(z)$.

The first and second terms of the equation appear close to the origin of $z$, where the optical paths of the probe and reference beams match each other.
On the other hand, the third term appears at around $z=z_d$, as indicated by convolution with delta-function $\delta(z-z_d)$.

Observing the third term, it is evident that the shape of this term is the sample structure $\alpha(z)$ convolved with the coherence function $\Gamma(z)$.
In other words, this term provides us the sample structure blurred by a point spread function $\Gamma(z)$.
Namely, this term is the OCT signal that we wish to measure.
Note that Eq.\@ (\ref{eq:Idz2}) represents a numerical signal obtained by the process described above.
Namely, the sample structure along the depth has been obtained as numerical data.

The fourth term is the mirrored complex conjugate of the third term (OCT signal), and is frequently denoted as a mirror signal or a coherent artifact.
Since the OCT signal and the mirror signal carry identical information, one can arbitrarily choose one of the signals.
In addition, the four signals are well separated if $\Delta z$ is properly selected.
Hence, it is easy to separate the OCT signal from the other terms.

\begin{figure}
	\centering\includegraphics{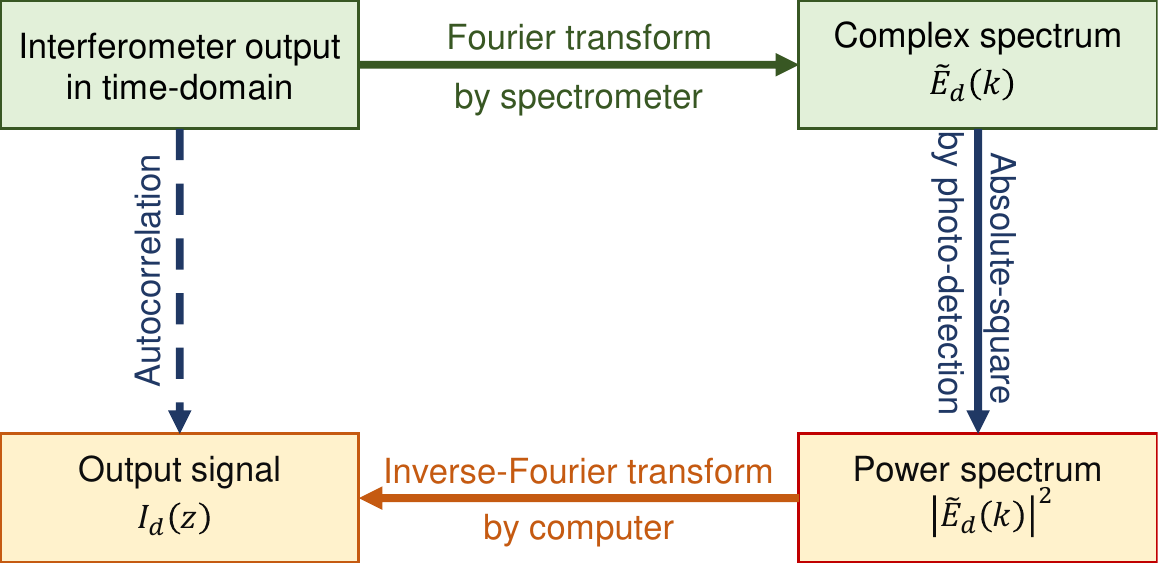}
	\caption{Processing flow of OCT signal reconstruction.
	The output from the interferometer in the time-domain (top left) is physically Fourier transformed into its time spectrum (top right) by a spectrometer.
	The spectrum is converted into the power spectrum (bottom right) through the intensity-sensitive photo-detection.
	The power spectrum is digitized and numerically inversely Fourier transformed by a computer and the output signal (bottom left) is formed. 
	Since the flow from the interferometer output to the output signal is equivalent to the flow of Wiener-Khinchin's theorem, the output signal becomes the autocorrelation function of the interferometer output.
}
	\label{fig:procSchem}
\end{figure}
This process is intuitively depicted in Fig.\@ \ref{fig:procSchem}.
The process flow starts with the complex temporal output from the interferometer, in which the sample structure is encoded in a time-of-flight manner in the probe beam.
The subsequent process extracts the sample structure from this signal.
First, the signal is physically Fourier transformed by the spectrometer.
The spectrometer first yields a joint complex spectrum of the probe and reference beams that is represented by Eq.\@ (\ref{eq:edk}).
The joint power spectrum is then converted into its squared power during the detection process.
Hence, from the line sensor we obtain the joint power spectrum of the probe and reference beams.
This joint power spectrum is digitized and numerically inverse-Fourier transformed into $I_d(z)$.

This process flow can be summarized as follows: an input signal is first Fourier transformed, squared, and then inverse-Fourier transformed.
Summarized this way, the process flow of SD-OCT is found to be equivalent to the calculation flow of the Wiener-Khinchin's theorem.
This suggests that the final output of this flow $I_d(z)$ is the autocorrelation of the input signal.
Since the input signal consists of the probe and reference beams, the output signal includes not only the autocorrelations of the probe beam and the reference beam, but also the crosscorrelation between them.
These cross-correlation terms are represented by the third and fourth terms of Eq.\@ (\ref{eq:Idz2}), i.e., the OCT and mirror signals.

The third and final part of the OCT reconstruction is logarithmic compression.
This process is needed because of the sensitivity of OCT, and hence the SNR of an OCT image is significantly higher than the color resolution of an OCT monitor (computer display) or the contrast sensitivity of a human eye (observer).
For instance, an SD-OCT system with a standard configuration possesses a sensitivity of around 98 dB \cite{Wojtkowski2004OpEx, Cense2004OpEx}.
Assuming the reflectivity of the eye is -50 dB, the SNR of an OCT image is around 48 dB.
On the other hand, the typical contrast resolution of a computer display is 256 gray scales, which is only around 24 dB.

To properly display the OCT signal on a computer screen, the OCT signal is commonly compressed into logarithmic space using $10\log\left|I_d(z)\right|^2$.
Here, the base of the logarithm is 10, and $I_d(z)$ is absolute-squared in order to convert it to signal energy.
Because $\left|I_d(z)\right|^2$ is an energy, $10\log\left|I_d(z)\right|^2$ represents the OCT signal in dB scale.
This definition of dB scale is conventionally utilized for OCT.

\subsection{SD-OCT and SS-OCT}
\label{sec:SdVsSs}
As mentioned in previous sections, there are two implementations of FD-OCT; SD-OCT and SS-OCT \cite{Yun2003OFDI, Yasuno2005OpEx, Huber2005OpEx}.
SD-OCT uses a broadband light source and the spectrally resolved detection is achieved by a high speed spectrometer.
On the other hand, SS-OCT uses a non-spectrally resolved point detector.
The spectral resolution of SS-OCT is achieved by a high-speed wavelength scanning light source, a laser with a high-speed wavelength tuner in its cavity \cite{Klein2013BOE, Yun2003OFDI, Huber2005OpEx, Oh2005OL}.
Because of the wavelength scanning, a spectrally resolved interferogram is obtained as a time-resolved signal measured by the point photodetector.
SS-OCT is regarded as a monochrometer-version of FD-OCT. 

The decision to use SD-OCT or SS-OCT is mainly based on the probe wavelength.
As described in Section \ref{sec:principle}, SS-OCT uses a line sensor to detect the spectral interferogram.
This line sensor is typically a silicon device that is only sensitive to wavelengths shorter than 1 \um.
Hence, FD-OCT with a short probe-wavelength is commonly implemented as SD-OCT.
It should be noted that longer-wavelength FD-OCT can also be implemented as SD-OCT \cite{Makita2008OpEx, Yun2003OpEx_SD} if an InGaAs line sensor is used.
However, the relatively high price of InGaAs line sensors has impeded the popularization of SD-OCT with long-wavelength probes.

FD-OCT using a long probe wavelength such as 1060 or 1310 nm is commonly implemented as SS-OCT.
One potential drawback of SS-OCT is its wide detection bandwidth.
For instance, if the acquisition time of a single spectral interferogram is $\tau$, the detection bandwidth of SD-OCT is $1/\tau$.
On the other hand, the bandwidth of SS-OCT is $N/\tau$, where $N$ is the number of sampling points in a single spectral interferogram.
Here we have assumed a 100\% detection cycle duty.
In a typical FD-OCT implementation, $N$ is about 1,000-2,000.
Namely, the detection bandwidth is 1,000-2,000 times wider in SS-OCT than in SD-OCT.
This wider detection bandwidth results in fewer rejections of relative intensity noise (RIN) of the light source.
Hence, SS-OCT frequently operates at the RIN limit, while SD-OCT easily operates close to the shot-noise limit (see also Section \ref{sec:snrSens}).
This drawback of SS-OCT can be partially compensated for by using a balanced photodetection scheme.
Because of the RIN rejection capability of balanced detection, recent SS-OCT systems can operate close to the shot-noise limit, i.e., only a few dB below it.
This difference in the detection scheme, i.e., balanced or non-balanced, results in different interferometer designs for SD- and SS-OCT.

\subsection{Interferometer design of FD-OCT}
\label{sec:interferometer}
FD-OCT interferometers can be classified into balanced or non-balanced.
The non-balanced interferometer is commonly configured as a Michelson interferometer and is used for SD-OCT.
On the other hand, the balanced interferometer is used for SS-OCT and can be implemented as either a Mach-Zehnder or Michelson interferometer.

\subsubsection{Non-balanced Michelson interferometer}
\begin{figure}
	\centering\includegraphics{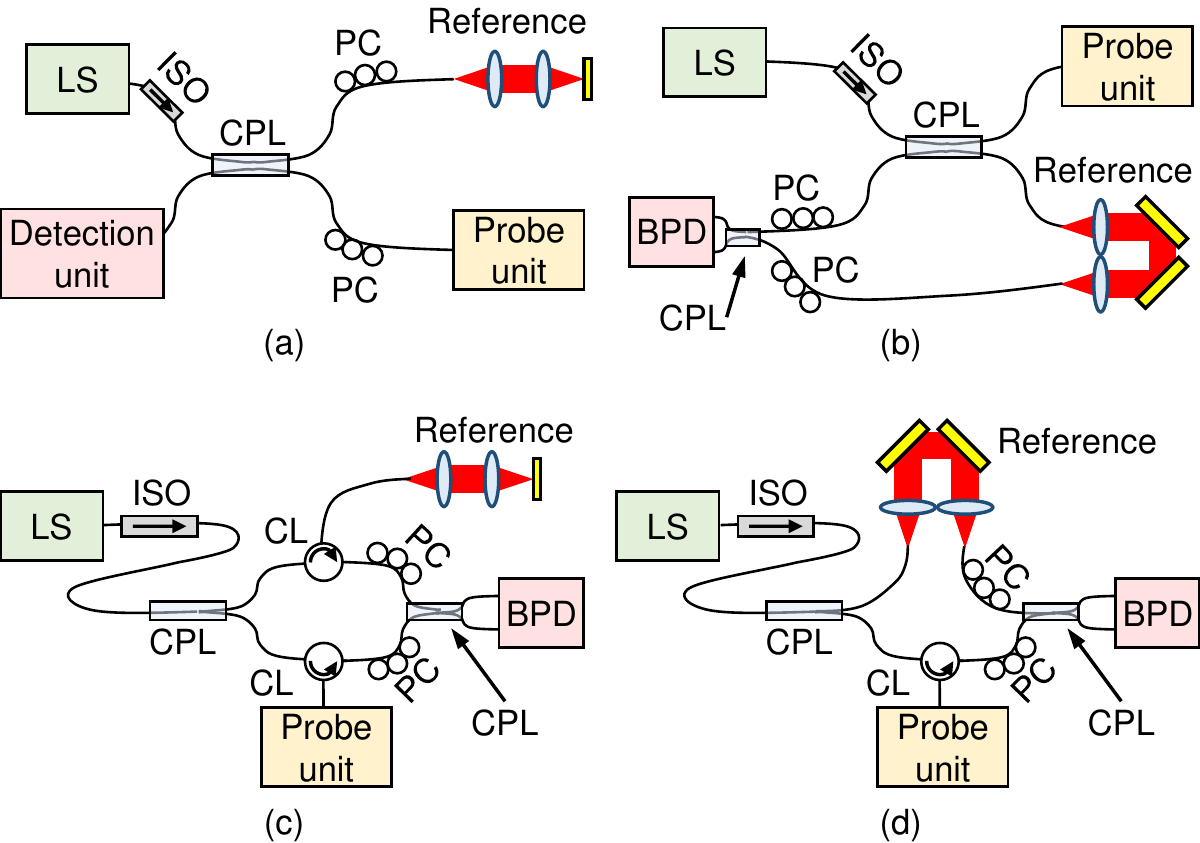}
	\caption{Examples of interferometer configurations: Non-balanced Michelson interferometer (a), balanced Michelson interferometer (b), and balanced Mach-Zehnder interferometers (c) and (d).
	Here, LS is a light source, ISO is an isolator, CPL is a coupler, CL is a circulator, PC is a polarization controller, and BPD is a balanced photo-detector.
	The non-balanced Michelson interferometer is suitable for SD-OCT, while the balanced interferometers are suitable for SS-OCT.	
	The balanced Michelson interferometer is suitable for the wavelength fo which a good circulator is not available.
}
	\label{fig:interferometer}
\end{figure}
As discussed in Section \ref{sec:SdVsSs}, SD-OCT does not require balanced photodetection because of its narrow detection bandwidth.
Hence, SD-OCT is typically implemented with a non-balanced Michelson interferometer.
An example of a non-balanced Michelson interferometer is depicted in Fig.\@ \ref{fig:interferometer}(a).

The advantage of the non-balanced Michelson interferometer is its simplicity.
The minimally required fiber component for this type of interferometer is a coupler.
An isolator and several polarization controllers are also frequently used.
The former is utilized to reject back reflection light to the light source and consequently avoid photodamaging it.
The latter is to maximize the OCT signal, achieved when the polarization states of the probe and reference beams are identical at the detector.

A disadvantage of the Michelson interferometer is the inevitable loss of the probe beam.
Because the total throughput of the probe beam as well as the collection rate of the back scattered probe beam are defined by the splitting ratio of the coupler, the throughput cannot be 100\%.
The maximum total throughput is obtained with a 50/50 coupler and is 25\%.

However, for eye imaging, the probe collection rate is more important than the total throughput.
The probe power on the eye is limited by a safety standard.
Thus, the collection rate is the primary factor that limits the sensitivity.
One typical example of the coupler for retinal imaging is a 20/80 coupler.
The 20\% portion of the beam is utilized as a probe beam, and the 80\% portion of the back scattered beam from the sample is collected for imaging.
Although the total throughput of this configuration is only 16\%, this configuration provides better sensitivity than the 50/50 coupler when the probe power on the eye reaches the safety limit.

A non-balanced Michelson interferometer is typically utilized for SD-OCT at any wavelength.

\subsubsection{Balanced Mach-Zehnder interferometer}
Typical configurations of the balanced Mach-Zehnder interferometer are depicted in Figs.\@ \ref{fig:interferometer}(c) and (d).
In these configurations, the light is first split into probe and reference beams by a coupler.
These beams are reflected and back scattered by a reference mirror and a sample and then the recombined by another coupler.

The biggest advantage of the Mach-Zehnder interferometer is its high probe beam throughput.
Because it uses a circulator, all the optical power of the back scattered probe beam contributes to image formation.
Since the OCT sensitivity is proportional to the probe beam power at the detector \cite{Leightgeb2003OpEx, deBore2003OL, Choma2003OpEx}, this advantage directly results in high system sensitivity.

A possible demerit of the Mach-Zehnder interferometer is the circulator.
Although a circulator is a well-established device for 1310-nm wavelengths, it is not well-developed for 1050 nm.
Hence, the circulator for 1050 nm is more expensive than that for 1310 nm, and its performance is not excellent.
For example, circulators are known to have polarization mode dispersion (PMD).
Although, the PMD of the circulator is negligible for standard 1310-nm OCT, it significantly affects a polarization sensitive extension of OCT known as polarization sensitive OCT (PS-OCT) \cite{EZhang2011OpEx, EZhang2013OpEx, Villiger2013OL}.
Because the circulator performance is generally worse at 1050 nm, circulator PMD is not negligible, even for a standard non-polarization-sensitive OCT at 1050 nm.

In summary, the balanced Mach-Zehnder interferometer is suitable for 1310-nm SS-OCT.

\subsubsection{Balanced Michelson interferometer}
Another interferometer for SS-OCT is the balanced Michelson interferometer.
A typical configuration is depicted in Fig.\@ \ref{fig:interferometer}(b).

The main advantage of this type of interferometer is balanced photodetection without the need for circulators.
The balanced detection scheme means that this interferometer is suitable for SS-OCT.
In addition, the absence of a circulator makes it suitable for 1050-nm wavelengths.
Hence, this interferometer is frequently utilized for 1050-nm SS-OCT \cite{Rollins1998OpEx, Braaf2011OpEx}.

Potential drawbacks of this interferometer are its limited total throughput and limited collection ratio of the probe beam.
These limitations are identical to that of a non-balanced Michelson interferometer.
Hence, a similar selection strategy for the coupler splitting ratio can be utilized to optimize the interferometer for eye imaging.

\subsection{Wavelength selection}
\label{sec:wavelength}
\begin{figure}
	\centering\includegraphics{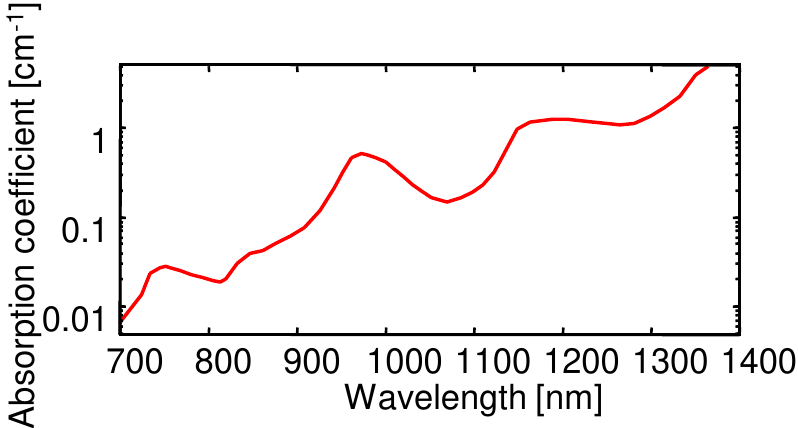}
	\caption{Absorption coefficient of water in the NIR region.
	In general, the water has higher absorption for longer wavelength. However, there are local minimum of the absorption at around 830 nm and 1050 nm.
	This plot was created by using the data presented in Ref.\@ \citen{Hale1973AO}.}
	\label{fig:waterAbsorption}
\end{figure}
Because water is the main light absorption source in living tissue, the wavelength of the OCT probe beam is selected to match one of local minima of water absorption.
Figure \ref{fig:waterAbsorption} shows the light absorption coefficients of water in the NIR region, where the plot was created using a dataset from Ref.\@ \citen{Hale1973AO}.
In this absorption curve, there are three local absorption minimums at approximately 830, 1050, and 1300 nm.
These three wavelengths are the most common OCT probing wavelengths.

The selection of one of these three wavelengths is roughly based on two general properties of light.
The first property is that the shorter wavelength is less absorbed by water.
This property leads to the first rule of penetration; the shorter wavelength probe penetrates the tissue more deeply.
The second property is that, in general, longer wavelengths are less scattered by the tissue.
The first rule is therefore in conflict with that of the second, namely that the longer wavelength has higher penetration.
In practice, tissue penetration is defined to balance these two rules.
Hence, the wavelength for the best penetration greatly depends on the scattering and absorption properties of the sample to be measured.
The probe wavelength is then selected mainly to best penetrate a specific sample.

\subsubsection{The 830-nm band}
In the NIR region, water has the lowest absorption at 830 nm.
On the other hand, this wavelength is the most highly scattered by the tissues.
Hence, the 830-nm probe is optimal for tissue that has fewer scatterers and comprises large amounts of water.
This suggests that the retina is the perfect target for 830 nm.

First, the retina possesses relatively fewer scatterers.
The photoreceptor cells that convert optical signals to electric signals are located at the outer (posterior) part of the retina.
Hence, when we see objects, the light that carries the visual information stimulates the photoreceptor cells after passing through the inner part of the retina.
This suggests that the scattering of the inner retina is low; otherwise the human visual function would be poor.
Because of this low scattering, the 830-nm probe can achieve relatively high penetration into the retina, even though it is more highly affected by scattering than other, longer wavelengths.

Another characteristic condition of retinal imaging concerns the optical media of the eye.
The retina is positioned beneath the eye optics that includes the cornea, aqueous humor, crystalline lens, and vitreous that forms a total length of around 24 mm, i.e., a 48-mm round trip for the probe beam.
These tissues mainly consist of water and have nearly no scattering.
Hence, for retinal imaging, the probe-beam should be less absorptive by water.

These two conditions make the 830-nm probe beam particularly suitable for retinal imaging.
It is also noteworthy that OCT with an 830-nm probe generally has better depth resolution than with other wavelengths.
This is because the depth resolution of OCT is proportional to the square of the center wavelength (see Section \ref{sec:resAndDisp}).
Because a retina consists of a fine-layered structure, this high resolution provides particular advantages for clinical examination.
Hence, almost all commercial retinal OCT is implemented with an 830-nm probe.

A Ti:Sapphire mode-locked laser and an SLD are two of the most common light sources for 830-nm OCT.
In practice, the former has a center wavelength of about 800 nm, while that of the latter is between 830–840 nm.
For clinical retinal OCT, SLD is more commonly utilized than the Ti:Sapphire laser because of its stability, cost-effectiveness, and compactness.
A typical SLD utilized for retinal SD-OCT has a center wavelength of 840 nm and a bandwidth of 50 nm, providing a depth resolution of around 5 \um in tissue.

\subsubsection{The 1050-nm band}
The 1050-nm band is an emerging wavelength band for retinal imaging.
Because water absorption has a second minimum at 1050 nm, this wavelength can pass through the ocular optical medium.
In addition, this relatively long wavelength suppresses scattering by the retinal tissue and provides higher penetration than the 830-nm probe.
This high penetration enables investigation of the deep posterior eye, i.e., the choroid.

Retinal imaging at the 1050 nm band was demonstrated with TD-OCT on \exvivo porcine eyes in 2003 \cite{Povazay2003OpEx} and for \invivo human eyes in 2004 \cite{Unterhuber2005OpEx}.
Since 2006, the 1050-nm band has been commonly utilized for retinal SS-OCT \cite{Lee2006OpEx, Yasuno2007OpEx, Srinivasan2008IOVS}.
Retinal SD-OCT at 1050 nm has also been demonstrated \cite{Makita2008OpEx, Povazay2007JBO, Puvanathasan2008OL}.
Although only one model of retinal 1050-nm OCT is commercially available as of 2014, several other commercial prototypes have also been demonstrated.
Hence, 1050 nm is expected to be the next generation of clinical retinal OCT, replacing 840-nm SD-OCT.

In general, optical and fiber components for 1050-nm beams are more expensive than other wavelengths and of relatively low quality.
However, recent developments in Ytterbium fiber lasers with a wavelength of 1060 nm have coincidentally driven the development of cost-effective and high-quality optical and fiber components at this wavelength.
The availability of these components further motivates the use of 1050 nm for the next generation of clinical retinal OCT.

\subsubsection{The 1310-nm band}
The 1310-nm band is a telecom wavelength band, and hence, the optical and fiber components at this wavelength are generally cost-effective and high quality.
Hence, this band has been widely utilized for OCT.

Compared with the other OCT wavelengths, 1310-nm light is minimally scattered by tissue.
Hence, it has very high tissue penetration.
As a result, this wavelength band is the most common wavelength for several clinical applications such as dermatology, cardiology, gastrointestinal imaging, and so on.
For ophthalmology, this wavelength band is utilized for anterior eye imaging \cite{Yasuno2005OpEx, Radhakrishnan2001ArchOphthalmol}.
On the other hand, 1310-nm light cannot be utilized for retinal OCT because it is highly absorbed by water.

\subsection{Resolution and dispersion}
\label{sec:resAndDisp} 
\subsubsection{Transversal resolution}
Since OCT is a point-scanning imaging modality and the signal intensity is proportional to the field amplitude instead of the squared power of the probe, the OCT transversal resolution is defined as the spot size on the sample.
Namely, assuming a Gaussian incident beam with diameter $d$, the transversal resolution $\Delta x$ becomes
\begin{equation}
	\label{eq:xres}
	\Delta x = \frac{4\lambda_c}{\pi} \left(\frac{f}{d}\right),
\end{equation}
where $\lambda_c$ is the center wavelength of a probe beam and $f$ is the focal length of the focusing lens \cite{OctHandbook}.
This equation suggests that the transversal resolution can be improved by increasing the incident beam diameter or decreasing the focal length.

However, the depth of focus, computed as $\pi \Delta x^2 / 2\lambda$, decreases as the transversal resolution increases.
Hence, the transversal resolution should be properly selected to maintain a sufficient depth of focus.

For retinal imaging, the focal length is that of the eye optics and hence is not selectable.
On the other hand, the beam diameter can be increased to achieve better transversal resolution.
One specific issue for retinal imaging is eye aberrations.
Because the human eye has a significant amount of aberrations, over-increasing the beam diameter sometimes worsens the transversal resolution.
Typically, the best achievable transversal resolution of retinal OCT is around 20 \um.

It is worth noting that high transversal resolution, as high as transform-limited resolution, can be achieved by a retinal OCT equipped with adaptive optics \cite{Zawadzki2005OpEx, Felberer2014BOE}.

\subsubsection*{Tutorial Exercise}
Calculate the transversal resolution of the following example of a typical OCT configuration.
The focal length of the lens is 40 mm, the center wavelength is 1.3 \um, and the beam diameter is 1.8 mm. 

\subsubsection*{Solution}
By substituting $f=40\times10^{-3}$  m, $\lambda_c=1.3\times10^{-6}$  m, and $d = 1.8 \times 10^{-3}$ m into Eq.\@ (\ref{eq:xres}), the transversal resolution is obtained as $\Delta x \simeq 3.7$ \um.

\subsubsection{Depth resolution}
Equation (\ref{eq:Idz2}) suggests that the OCT axial point spread function (PSF) is defined by the coherence function $\Gamma(z)$ which is a Fourier transform of the intensity spectrum of the light source.
Hence the depth resolution of OCT is defined by the width of the coherence function.

Assuming a Gaussian intensity spectrum with a full width at half maximum (FWHM) of $\Delta k$ in wavenumber, the full width of -6 dB of the maximum of $\Gamma(z)$ is obtained as $\Delta z = 4 \ln 2 / \Delta k$.
This equation can be approximated using the center wavelength of the probe beam $\lambda_c$ and the FWHM of wavelength of $\Delta \lambda$ as
\begin{equation}
	\label{eq:resz}
	\Delta z = \frac{2 \ln 2}{\pi} \frac{\lambda_c^2}{\Delta \lambda}.
\end{equation}
This $\Delta z$ is commonly utilized as the depth resolution of OCT.

Note that this resolution was derived assuming a Gaussian spectrum.
Recent wide-band SLDs frequently have non-Gaussian spectra, and this equation provides only a rough measure of depth resolution for such light sources.

\subsubsection*{Tutorial Exercise}
The specifications of an example of commonly utilized SLD light sources for SD-OCT are following.
Calculate the in-air and in-tissue depth resolutions of OCT utilizing this lightsource.
The lightsource has a Gaussian-shaped spectrum with FWHM of 55 nm and its center wavelength is 840 nm.
The refractive index of the air is 1.0 and that of tissue can be assumed to be 1.38.

\subsubsection*{Solution}
By substituting $\lambda = 840 \times 10^{-9}$ m and $\Delta \lambda = 55 \times 10^{-9}$ m into Eq.\@ (\ref{eq:resz}), the depth resolution is obtained as $\Delta z \simeq 5.7$ \um in air.
By dividing the in-air resolution by the refractive index of the tissue (1.38), the in-tissue resolution is obtained as $\Delta z \simeq 4.1$ \um.

\subsubsection{Dispersion and dispersion-correction}
Recall that Eq.\@ (\ref{eq:Idz2}) was derived by assuming the probe and reference arms have the same high order dispersion.
To consider the dispersion effect in OCT imaging, we need to remove this assumption.

The OCT equation that does not make this assumption [Eq.\@ \ref{eq:idz}] shows that the OCT signal is blurred by the function $\Gamma(z)*\ft{\tilde{\gamma_p}(k) \tilde{\gamma_r}\conj(k)}$.
To consider different high order dispersions, here we let $\tilde{\gamma_p}(k) = \sexp{i \phi_p(k)}\sexp{-ikz_p}$ and $\tilde{\gamma_r}(k) = \sexp{i\phi_r(k)}\sexp{-ikz_r}$, where $\phi_p(k)$ and $\phi_r(k)$ represent high order dispersions in the probe and reference arms, respectively.
The above function then becomes $\Gamma(z)*\ift{\sexp{i\phi_d(k)}}*\delta(z-z_d)$, where $\phi_d(k) = \phi_p(k)-\phi_r(k) $ and represents an unbalanced dispersion between the two arms.
Ignoring the shifting, the axial PSF under unbalanced dispersion then becomes
\begin{equation}
	\label{eq:PSF}
	\mathrm{PSF} = \Gamma(z)*\ift{\sexp{i\phi_d (k)}}.
\end{equation}
Namely, the PSF is additionally broadened by the unbalanced dispersion.
Another remarkable point is that the PSF under dispersion is asymmetric, while PSF without dispersion is symmetric, because the inverse-Fourier transform of an arbitrary function becomes asymmetric unless the function is a real function.
This fact can be utilized to identify the cause of unexpected PSF broadening when building an OCT system.

Dispersion unbalance comes not only from the OCT system but also from the sample itself.
Particularly for retinal imaging, the ocular media causes dispersion unbalance.
In addition, the ocular dispersion varies subject by subject.
Hence, dispersion correction is essential for retinal imaging.

Unbalanced dispersion can be roughly cancelled by designing the OCT optics to have minimally unbalanced dispersion.
However, this hardware cancellation cannot cancel the residual dispersion that varies among subjects.
This residual dispersion is commonly canceled using numerical methods.

Using the same $\phi_p(k)$ and $\phi_r(k)$ as in Eq.\@ (\ref{eq:PSF}), the third term of Eq.\@ (\ref{eq:idk}), i.e., the component of spectral interferogram corresponding to the OCT signal, becomes $\sqrt{p'_p p'_r}\left|\tilde{s}(k)\right|^2 \tilde{\alpha(k)}\sexp{i \phi_d(k)}\sexp{-ikz_d}$, where $\sexp{-i\phi_d(k)}$ is the unbalanced dispersion.
Because the interferogram has been digitized, the unbalanced dispersion can be canceled by numerically multiplying the interferogram by the complex conjugate of $\sexp{-i\phi_d(k)}$ , i.e., $\sexp{i\phi_d(k)}$ , called the counter dispersion.

Unfortunately, $\sexp{-i\phi_d(k)}$ is normally unknown.
Hence, iterative optimization algorithms are utilized to cancel the unbalanced dispersion.
In such algorithms, an estimated counter dispersion is applied to the spectral interferogram before the Fourier transform.
Subsequently, a sharpness metric is computed from a reconstructed OCT image.
The counter dispersion is then iteratively optimized to maximize the sharpness metric.
Several types of sharpness metrics have been utilized for retinal imaging, including the number of high intensity pixels \cite{Wojtkowski2004OpEx} and image information entropy \cite{Yasuno2007OpEx}.

\subsection{SNR and sensitivity}
\label{sec:snrSens}
\subsubsection{Definition and measurement of SNR and sensitivity}
Sensitivity and SNR are commonly utilized to evaluate the imaging performance of OCT.
The sensitivity is defined as the maximum measurable attenuation of the probe beam.
Thus, sensitivity is a specification of the system.
On the other hand, SNR is the ratio of the signal energy (at a certain point in the image) to the noise energy and hence is a specification of an image.
Although the sensitivity and SNR are specifications of two different things, they are tightly coupled.
Namely, the SNR is proportional to the sensitivity when a same sample is measured at a same position.

The sensitivity of an OCT system is commonly measured in the following manner.
An OCT signal is acquired using a perfectly reflecting mirror with a neutral density (ND) filter placed before the mirror as a sample.
This ND filter mimics the attenuation of a target tissue.
In the case of retinal imaging, an ND filter with an optical density of 2.5–3.0 is utilized.
This gives a round-trip attenuation of -50 to -60 dB and mimics the reflectivity of the eye.
Absorption-type ND filters are not optimal for this purpose because they sometimes have a large dispersion.
For this measurement, transversal scanning should not be performed to avoid inessential signal attenuation.
The signal intensity at the mirror is then determined from the OCT image.
Multiple A-lines are then obtained with the mirror sample removed.
A series of complex OCT signals are taken at the position where the mirror sample was located, and the variances of the real and imaginary parts of the OCT signal, $\sigma_r^2$  and $\sigma_i^2$, are obtained.
The noise energy can then be determined as a summation of the variances, i.e., $\sigma_r^2 +\sigma_i^2$.
The SNR of the OCT image of the mirror is then determined as the ratio between the signal and noise energies.
Finally, the sensitivity is determined in dB scale by subtracting the SNR from the attenuation of the ND filter.
For example, if the attenuation of the ND filter is -60 dB and the mirror signal has an SNR of 40 dB, the sensitivity becomes -60 dB $-$ 40 dB = -100 dB.

According to its literal definition, the sensitivity is a negative value in dB.
However, it is conventionally presented as a positive value by omitting the negative sign.
We also use this convention in this chapter.
With this convention, the SNR becomes proportional to the sensitivity in linear scale, or a constantly biased value in dB.
Particularly for a sample with 100\% of reflectivity, the SNR is identical to the sensitivity.

\subsubsection{Noise source and sensitivity optimization}
There are three major noise sources in OCT.
The first one is shot noise that originates from statistical fluctuations in counting photons and/or electrons.
Because it is a fluctuation in counting, shot noise is proportional to the square-root of the optical power, and its energy is proportional to the optical power.
Namely, short noise energy $\sigma^2_{shot}$ is proportional to $p'_p + p'_r$.
In practical OCT imaging, we can assume $p'_r \gg p'_p$ and hence, $\sigma^2_{shot} \propto p'_r$.

The second source of noise is the RIN of the light source.
Since RIN is a fluctuation in light intensity, its energy $\sigma^2_{RIN}$ is proportional to $\left(p'_p + p'_r\right)^2$.
With the same assumptions as for shot-noise, we can say $\sigma^2_{RIN} \propto p'^2_r$.

The third source of noise is detection noise.
This is an aggregation of non-optical noises such as the thermal noise of the photodetector and electrical noise of a signal digitizer.
As evident from its definition, the energy of detection noise is independent to $p'_p$ and $p'_r$.
Here, detection noise energy is denoted as $\sigma^2_{det}$.

The energy (i.e., squared power) of the OCT signal $\epsilon_s$ is proportional to $p'_p p'_r$, as indicated by Eqs.\@ (\ref{eq:idz}) and (\ref{eq:Idz2}).
Using the energies of the signal and various noises, the SNR is expressed as
\begin{equation}
	\label{eq:snr}
	\mathrm{SNR} \equiv \frac{\epsilon_s}{\sigma^2_{shot} + \sigma^2_{RIN} + \sigma^2_{det}}
	= \frac{c_a p'_p p'_r}{c_b p'_r + c_c p'^2_r + \sigma^2_{det}},
\end{equation}

where $c_a$, $c_b$,and $c_c$ are proportional constants that are defined by hardware specifications.
Assuming a mirror sample with 100\% reflectivity, this equation also represents the sensitivity.

Although we omit discussion of the practical values of the proportional constants, this equation provides some important insights into OCT sensitivity.
Most importantly, OCT sensitivity is proportional to probe power.
Therefore, it is a common strategy to increase the probe power in order to improve sensitivity.
Second, sensitivity is a function of the power of a reference beam.
From this perspective, the OCT operation can be classified into three regimes: shot-noise limited, RIN limited, and detection-noise limited.
In each regime, the sensitivity has a different dependency on the reference power.

The regime where $\sigma^2_{shot} \gg \sigma^2_{RIN}, \sigma^2_{det}$ is the shot-noise limited regime.
In this regime, the sensitivity is approximated by $\propto \left(c_a / c_b\right) p'_p$ and is evidently independent from the reference power.
Similarly, in the RIN limited regime, the sensitivity is proportional to $\left(c_a / c_c\right) \left(p'_p/p'_r\right)$.
Namely, the sensitivity is inversely proportional to the reference power.
In the detection-noise limited regime, the sensitivity is proportional to $\left(c_a / \sigma^2_{det}\right) \left(p'_p p'_r\right)$ and hence proportional to the reference power.

\begin{figure}
	\centering\includegraphics{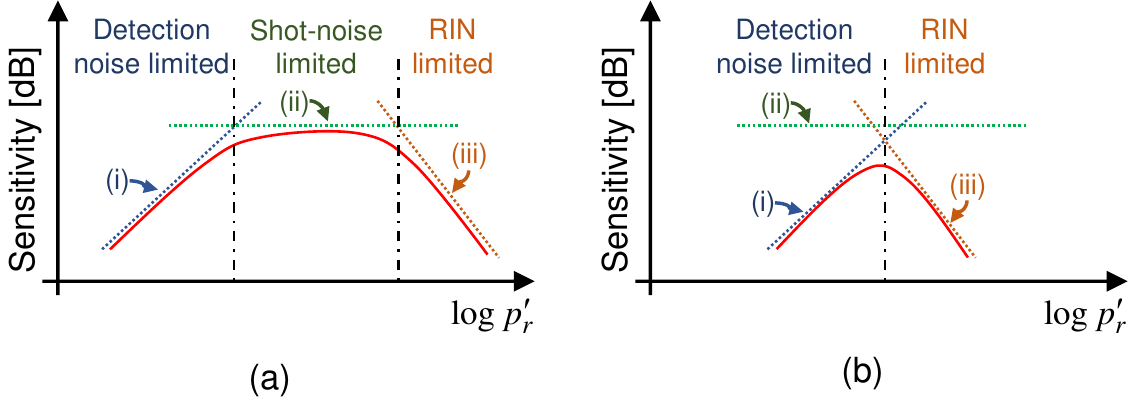}
	\caption{Examples of OCT sensitivity curves, where the sensitivity is plotted in a dB-scale as a function of logarithm of the power of a reference beam on a detector.
		The solid curves are the sensitivity curve, the dashed lines (i)-(iii) are the sensitivity curves assuming only detection noise, shot-noise, or RIN exists, respectively.
		The OCT can operate as a shot-noise limited regime in (a), but cannot in (b). 
	}
	\label{fig:senscurve}
\end{figure}
These facts are summarized in Fig.\@ \ref{fig:senscurve}(a), where the logarithmic sensitivity is plotted as a function of the logarithm of the reference power on the detector.
The solid curve depicted in this figure is called a sensitivity curve.
The dashed curves (i)-(iii) are the sensitivity curves assuming only detection noise, shot-noise, or RIN exists, respectively.
The dashed-dotted lines divide the three noise-regimes.
As the reference power increases, first the sensitivity and subsequently the SNR increase.
In this regime, the sensitivity is dominated by the detection noise (the detection noise limited regime).
Once the sensitivity reaches the shot-noise limited regime, it becomes constant.
Finally, it begins to decrease in the RIN regime.
Hence, the OCT reference power should be optimized such that OCT works in the shot-noise limited regime.

Another situation regarding the sensitivity is depicted in Fig.\@ \ref{fig:senscurve}(b), where OCT can never operate at the shot-noise limit.
In this case, the reference power can be suboptimally optimized to have the maximum achievable sensitivity.
In addition, modifications of the OCT system should be considered, such as employing a light source with a lower RIN, using detector/detection electronics with lower detection noise, or both.

To roughly design an OCT system, shot-noise limited sensitivity is particularly important.
This is because shot-noise is inevitable noise and hence is the true theoretical limit of OCT sensitivity.
The shot-noise limited sensitivity of FD-OCT is given as \cite{Yun2003OFDI, deBore2003OL, Leightgeb2003OpEx}
\begin{equation}
	\label{eq:sens}
	\mathrm{Sensitivity} = \frac{\eta \tau}{h \nu_0}p'_p ,
\end{equation}

where $p'_p$ is the optical power of the probe on the detector when the sample is a mirror with 100\% reflectivity, $h$ is the Plank constant, $\nu_0$ is the center frequency of the light source, $\eta$ is the quantum efficiency of the photodetector, and $\tau$ is the integration time of a line sensor (for SD-OCT) or the inverse of the light source wavelength scanning frequency (for SS-OCT).

\subsubsection*{Tutorial Exercise}
Calculate the shot-noise limited sensitivity of an SD-OCT with the following specifications.
The light source is an SLD with 840-nm center wavelength.
The interferometer is a non-balanced Michelson interferometer.
A 30/70 coupler splits the beam into reference and probe arms and 30\% portion of the beam is lead to the probe arm.
The probe power on the sample is 750 $\mu$W.
The diffraction efficiency of a grating utilized in a spectrometer is 90\%.
The quantum efficiency of a line CCD sensor utilized in the spectrometer is 40\%.
The line rate of the sensor is 70 kHz and its duty is 100\%.

\subsubsection*{Solution}
The coupler leads 70\% portion of the probe beam reflected from a sample.
By considering this collection efficiency of the coupler and the diffraction efficiency of the grating, the probe power on the sensor becomes $p'_p = 750 \times 10^{-6} \times 0.7 \times 0.9$ W.
The exposure time of the sensor is obtained from the line rate and duty as $\tau = 1.0 / 70,000$ s.
The center frequency $\nu_0 = c/\lambda_c$ with $c$ as the speed of light.
And $\eta = 0.4$.

By substituting these values into Eq.\@ (\ref{eq:sens}), the shot-noise limited sensitivity is estimated as $1.14 \times 10^{10}$, which is 100.6 dB in logarithmic scale.

\section{OCT in ophthalmology}
\label{sec:ophthalmology}
In this section, we discuss about the utility of OCT in ophthalmology.
The modalities for ophthalmic diagnosis can be characterized by three factors.
The first factor is whether the modality is objective or subjective, the second factor is whether it is structural or functional, and the third factor is invasiveness.
For example, visual acuity and field tests are subjective, functional, and non-invasive examinations.
Similarly, angiographies, including fluorescein angiography and indocyanine green angiography (ICGA), are objective, functional, and invasive modalities.
Among other modalities, OCT is characterized as an objective, structural, and non-invasive modality.
Color fundus photography and fundus autofluorescence imaging are also objective, structural, and non-invasive, but these modalities only provide 2-D \enface images, while OCT can provide a 3-D volumetric tomography.

Thus far, we have not particularly differentiated OCT systems for posterior and anterior eye examination.
However, the optimal OCT is not the same for these two parts of the eye.
For example, the probe beam for posterior eye imaging should not be highly absorbed by the ocular media, as discussed in Section \ref{sec:wavelength}, while this condition is not significant for anterior imaging.
Additionally, the information needed for the diagnosis of posterior and anterior eye diseases differs, and hence, the optimal design of an anterior or posterior OCT also differs.
In this section, we independently discuss the features and utilities of posterior and anterior OCT.

\subsection{Posterior eye imaging}
\subsubsection{Structure of the posterior eye}
The posterior part of the eye consists of three major layers, the retina, choroid, and sclera, in anterior (inner) to posterior (outer) order.
The retina is the sensory part of the eye and consists of a further 10 layers, from the anterior to posterior: inner limiting membrane (ILM), nerve fiber layer (NFL), ganglion cell layer (GCL), inner plexiform layer (IPL), inner nuclear layer (INL), outer plexiform layer (OPL), outer nuclear layer (ONL), external limiting membrane (ELM), photo-receptors, retinal pigment epithelium (RPE), and Bruch's membrane.
The photo-receptor layer is further divided into two segments: inner and outer.
The choroid is a vascular-rich tissue beneath the retina that nourishes the posterior part of the retina.
The retina and the choroid are separated by the RPE, a melanin rich cellular monolayer, and Bruch's membrane.
The sclera is a further outer layer and provides mechanical support for the eye.
A narrow definition of the retina refers to only the retina, while the wider sense of the word refers to the retina and choroid.

\subsubsection{Posterior imaging at 830 nm}
\begin{figure}[tbh]
	\centering\includegraphics{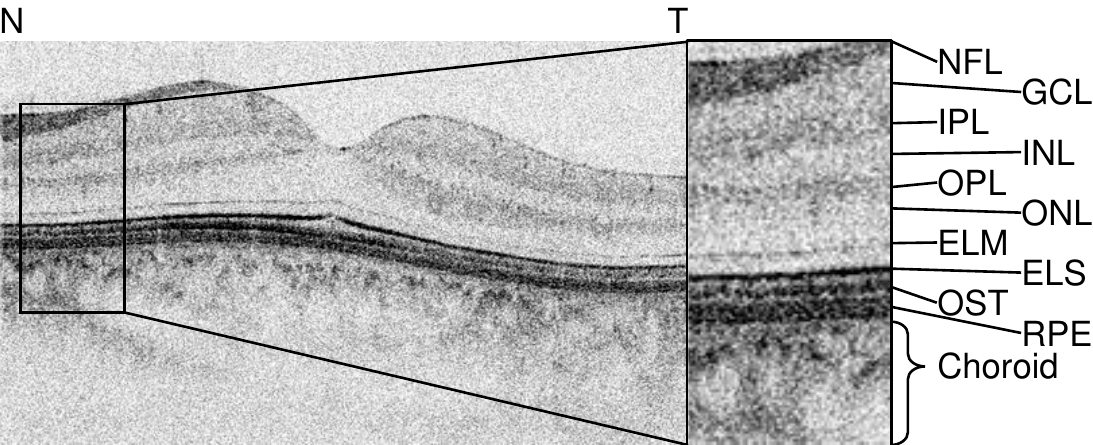}
	\caption{An example of an ultra-high resolution OCT of \invivo normal macula. Nasal and temporal directions are represented by N and T, respectively. NFL: nerve fiber layer, GCL: ganglion cell layer, IPL: inner plexiform layer, INP: inner nuclear layer, OPL: outer plexiform layer, ONL: outer nuclear layer, ELM: external limiting membrane, ELS: ellipsoid region, OST: the tip of photoreceptor outer segment, and RPE.}
	\label{fig:uhrRetina}
\end{figure}
Figure \ref{fig:uhrRetina} shows an example OCT cross section of an \invivo normal human macular, which is the visual center of the posterior eye.
The image was taken by an ultra-high-resolution SD-OCT \cite{Hong2007OpEx} at the 830-nm band (a center wavelength of 870 nm and a bandwidth of 170 nm) possessing a depth resolution of 2.9 \um in tissue.
The top and bottom of the image represents the anterior and posterior directions, respectively, and this image orientation is the convention for displaying posterior OCT images.
In this image, the OCT signals are displayed with an inverted-gray scale, i.e., black represents hyperscattering and white represents hyposcattering.
Because the ILM is too thin to be visualized, the topmost hyperreflective layer appearing in the OCT image is the NFL.
From the NFL to the posterior, the retinal layers shown are the GCL (hyposcattering), IPL (hyperscattering), INL (hyposcattering), OPL (hyperscattering), ONL (hyposcattering), and ELM (thin hyperscattering layer).
The strong hyperscattering layer beneath the ELM is the outermost part of the inner segment of the photoreceptor, called the ellipsoid region (ELS).
Although it is currently believed to a part of the inner segment, it was once believed to be the junction between the inner and outer segments of the photoreceptor, and hence is still sometimes denoted as the IS/OS.
Another hyperscattering line beneath the ELS is the outer tip of the outer segment of the photoreceptor (OST).
The final strong-scattering line beneath the OST is the RPE.
Bruch's membrane cannot be identified with the normal eye because it is closely attached to the RPE.
However, it sometimes appears as a faint hyperscattering line in some pathologic cases.
The ELS, OST, and RPE layers are sometimes denoted as the RPE complex for historical reasons.
Namely, these layers were hard to separate using early generations of posterior OCT because of their relatively low depth resolution.

The choroid can be roughly classified into three layers.
The inner-most layer is the choriocapillaris (CC) that is just beneath Bruch's membrane and appears as a thin hyperscattering layer in an OCT image.
Sattler's layer is located to the posterior of the CC and consists of medium-diameter blood vessels.
Finally, Haller's layer is located at the most outer part of the choroid and consists of large-diameter blood vessels.
The blood vessels in Sattler's and Haller's layers appear as hyposcattering regions in the OCT image.

The RPE and choroid contain certain amounts of melanin that highly absorb the probe beam at the 830-nm band \cite{Unterhuber2005OpEx, Yasuno2007OpEx}.
It limits the image penetration, and the image contrast of the choroid is very low at 830-nm.
Additionally, the sclera is almost invisible at this wavelength.

\begin{figure}[tbh]
	\centering\includegraphics[width=\textwidth]{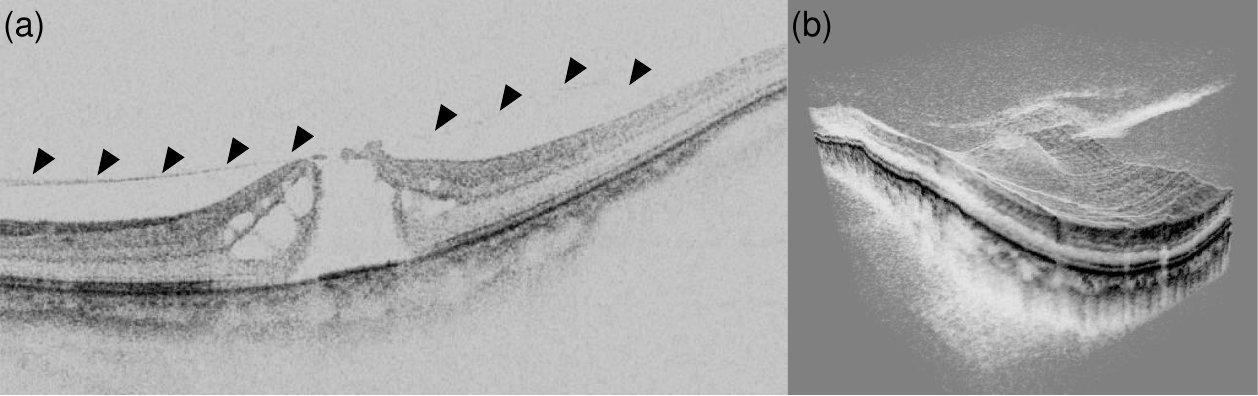}
	\caption{Cross-sectional 830-nm SD-OCT image of a macular hole (a) visualizes the detached vitreous membrane tracting the retina upward (arrow heads).
    The volume rendering of the same case (b) reveals the 3-D extension of the vitreous membrane.}
	\label{fig:macularhole}
\end{figure}
Despite the limited penetration, 830-nm SD-OCT is very widely used for posterior diagnosis in eye clinics.
An example of a clinical case using 830-nm SD-OCT is shown in Fig.\@ \ref{fig:macularhole}.
The case concerns a macular hole examined by an early ophthalmic SD-OCT prototype with a probe wavelength of 840 nm and an axial resolution of 4.5 \um in tissue (see Ref.\@ \citen{Makita2006OpEx} for details).
In the cross-sectional image (Fig.\@ \ref{fig:macularhole}(a)), the detachment of the retina from the RPE is clearly visible.
Additionally, the traction of the retina by detached vitreous is clearly visualized (arrowheads).
The volume rendering of the same case (Fig.\@ \ref{fig:macularhole}(b)) allows a more intuitive understanding of the pathology.

In addition to the qualitative observation of structural abnormalities in retinal pathology, OCT has been used to quantitate them.
The quantification is mainly done by measuring the thickness of the retinal layers.

The peripapillary NFL thickness is known as a good indicator of glaucoma.
Almost all clinical posterior OCT devices can automatically measure this thickness and provide a risk score for glaucoma.
Recently, the thicknesses of the inner retinal layers, i.e., NFL, GCL, and IPL, have been used as more sensitive measures of glaucoma \cite{Tan2008Ophthalmol}. 

\subsubsection{High-penetration posterior OCT at 1050-nm}
As discussed above, the early generations of posterior OCT that use a 830-nm probe do not provide high contrast images at the choroid.
This issue has been gradually resolved by 1050-nm OCT.

\begin{figure}[tbh]
	\centering\includegraphics{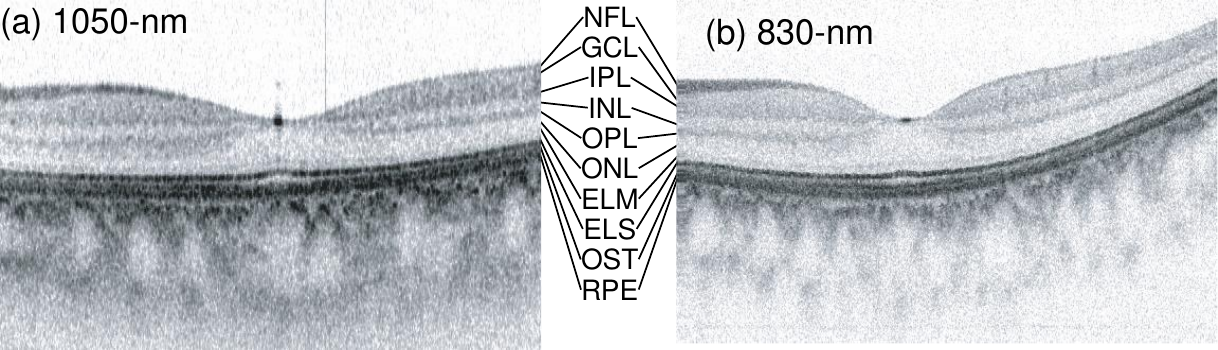}
	\caption{Retinal OCT images taken by 1050-nm SS-OCT (a) and 830-nm SD-OCT (b). The images are of the same macular region of a normal \invivo retina.
    Although the depth-resolution of the 1050-nm SS-OCT is lower than that of 830-nm SD-OCT, the retinal layers can be clearly identified.
    In addition, the former also visualizes the choroid with high contrast.
    The figures are reprinted from Ref.\@ \citen{Povazay2003OpEx} with modification.}
	\label{fig:sdVsSs}
\end{figure}
Figure \ref{fig:sdVsSs} shows a comparison of OCT images of a normal macula taken by 1050-nm SS-OCT [Fig.\@ \ref{fig:sdVsSs}(a)] and 830-nm SD-OCT [Fig.\@ \ref{fig:sdVsSs}(b)].
(The figures are modified and reprinted from Ref.\@ \citen{Povazay2003OpEx}.)
The 1050-nm SS-OCT and 830-nm SD-OCT respectively possess depth resolutions of 10.4 \um and 4.5 \um in tissue.
Although the depth resolution of 1050-nm OCT is lower, it clearly shows the retinal layers.
Additionally, it visualizes the choroid with high contrast.
In 1050-nm OCT image, the hyperscattering CC layer beneath the RPE and large hyposcattering structures that are large blood vessels in the Haller's fs layer are visible.
The interface between the choroid and sclera is also visible as a hyperscattering line. 

\begin{figure}[tbh]
	\centering\includegraphics{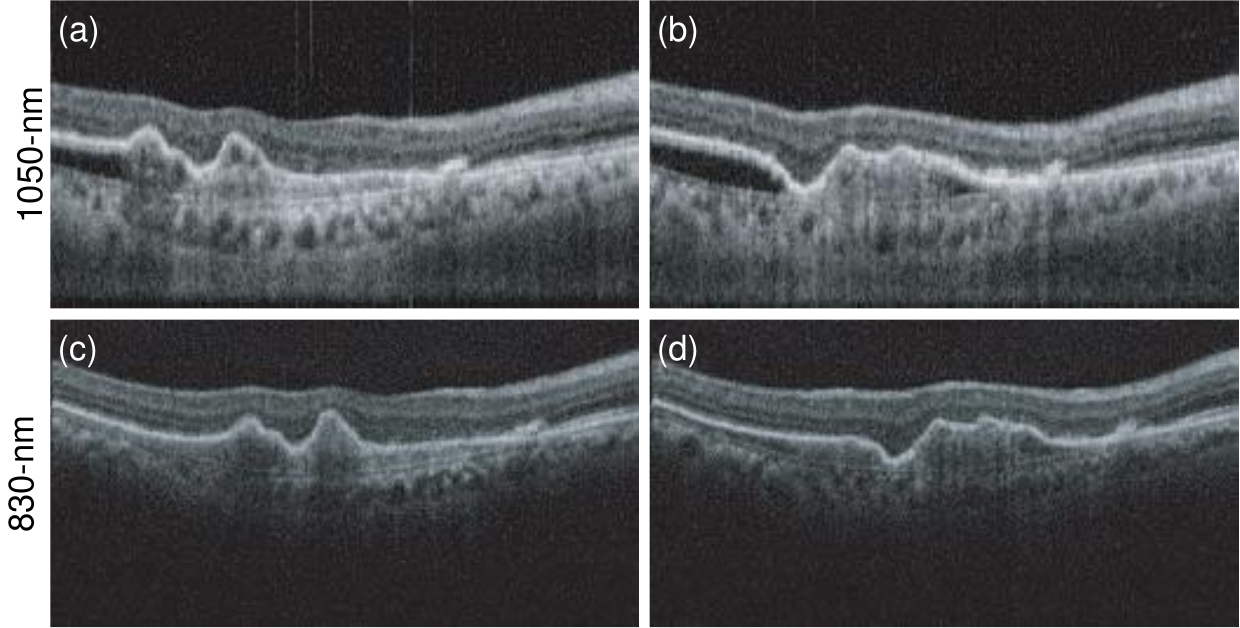}
	\caption{OCT images of polypoidal choroidal vasculopathy. (a) and (b) were taken by 1050-nm SS-OCT while (c) and (d) were taken by 830-nm SD-OCT.
    The 1050-nm SS-OCT images reveal the structures beneath the RPE detachment, where hyper-scattering tissues and void regions are visible.
    }
	\label{fig:830vs1060}
\end{figure}
OCT at 1050 nm can be used for both qualitative observation and quantitative examinations.
Figure \ref{fig:830vs1060} shows examples of OCT cross sections of a case of polypoidal choroidal vasculopathy.
The images were modified and reprinted from Ref.\@ \citen{Yasuno2009IOVS}.
The images are displayed in gray scale, i.e., bright pixels represent hyperscattering and dark pixels represent hyposcattering.
Figures \ref{fig:830vs1060}(a) and \ref{fig:830vs1060}(b) were obtained by 1050-nm SS-OCT, while Figs.\@ \ref{fig:830vs1060}(c) and \ref{fig:830vs1060}(d) were obtained by 830-nm SD-OCT.
Figures \ref{fig:830vs1060}(a) and \ref{fig:830vs1060}(c) and Figs.\@ \ref{fig:830vs1060}(b) and \ref{fig:830vs1060}(d) were taken at the same locations on the retina.
In all images, a severe detachment of the RPE is visible.
In the 830-nm images, the structure beneath the RPE detachment is unclear.
However, the 1050-nm images have imaged two types of regions beneath the RPE detachment: hyposcattering and hyperscattering regions.
A clinical study showed that the hyperscattering regions correspond to the polypoidal hyperfluorescence in ICGA \cite{Yasuno2009IOVS}.
Another remarkable feature of the 1050-nm images are their high penetration into the choroid and sclera.

Another clinical utility of 1050-nm OCT is its ability to quantify choroidal thickness because of the full-depth penetration into the choroid.
Because the choroid is known to be associated with several retinal diseases, a quantitative choroidal thickness measurement is expected to be a powerful tool for quantitative diagnosis.
Currently, the choroidal thickness has been found to be associated with several posterior eye diseases including age-related macular degeneration and polypoidal choroidal vasculopathy \cite{Chung2011Ophthalmol}, central serous chorioretinopathy \cite{Maruko2011Retina, Kuroda2013Retina}, and some whole-eye diseases including glaucoma \cite{Usui2012AJO} and myopia \cite{Fujiwara2009AJO, Ikuno2013CO}.

Although almost all commercial OCT systems for posterior eye imaging are 830-nm SD-OCT, a 1050-nm posterior OCT device is already commercially available and some other semi-commercial research prototypes were demonstrated in 2014.
These commercial and semi-commercial devices have shown high clinical utilities and are expected to be the next generation of clinical posterior OCT, replacing 830-nm SD-OCT.

\subsection{Anterior eye imaging}
The three targets of anterior eye imaging by OCT are the anterior chamber, tissues at the eye surface, and the cornea.
For the anterior chamber and the eye surface, 1310-nm OCT is mainly used because it has higher penetration than other wavelengths.
The purpose of corneal imaging can be further classified into visualization of tissue abnormalities and morphometric investigation.
Because of the two distinct purposes, both 830- and 1310-nm are used for this target. 

\subsubsection{Anterior eye chamber imaging}
When OCT imaging the anterior eye chamber, some parts are imaged in a transscleral manner, and hence, the anterior OCT should have a high penetration.
Conversely, the main interest of anterior OCT imaging is the gross morphology in the eye chamber.
Hence, the required resolution is lower than that of posterior OCT.
For these reasons, anterior eye chamber imaging is mainly performed with 1310-nm OCT.

Among several clinically significant targets in anterior chamber imaging, quantitative evaluation of anterior angle, the angle between the iris and cornea, is known to be useful for assessing the risk of angle closure glaucoma.
In more detail, the anterior eye chamber is filled with fluid called the aqueous humor.
The aqueous humor is generated at the ciliary body located posterior to the iris and drained out from the chamber through a channel located just anterior to the anterior angle called the trabecular meshwork.
If the trabecular meshwork is morphologically blocked by the iris, it causes an acute elevation of intraocular pressure that damages the optic nerve and results in angle closure glaucoma.
To prevent such an acute glaucoma attack, it is important to assess the risk of angle closure.
Quantitative assessment of the anterior angle using OCT aims to find eyes at high risk of glaucoma attack.

To quantitatively assess the anterior angle, a parameter that correctly reflects the morphology of the anterior eye chamber is required.
Some morphometric parameters originally developed for the analysis of the ultrasound images of anterior chamber, such as angle-opening distance at 500 \um (AOD500) \cite{Pavlin1992AJO} and angle recess area (ARA) \cite{Ishikawa2000COO} are commonly used for the morphological analysis of OCT.

\begin{figure}
	\centering\includegraphics{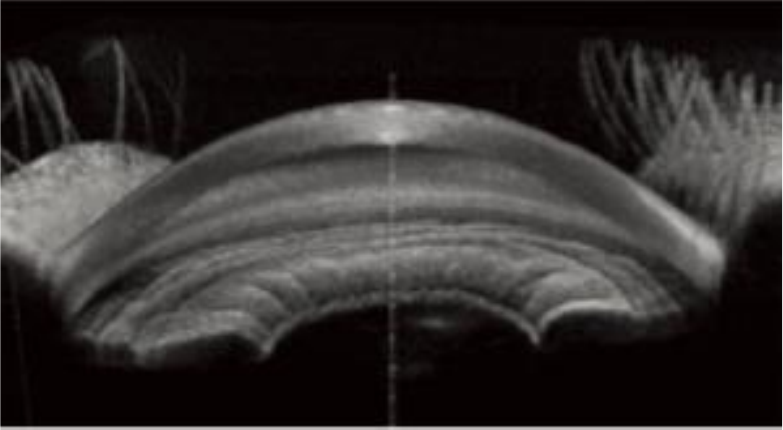}
	\caption{Example of whole circumference imaging of the anterior eye chamber.
    This visualization is similar to that of gonioscopy and enables intuitive understanding the the anterior angle structure.
    The image was provided from Tomey Corporation.}
	\label{fig:ac3d}
\end{figure}
The first generation of clinical anterior eye OCT was 1310-nm TD-OCT, and hence it provides only 2-D cross-sections of the anterior eye chamber.
However, the recent generation of anterior eye OCT is 1310-nm SS-OCT and provides a full 3-D morphology of the anterior eye.
Because of this 3-D capability, anterior SS-OCT provides whole circumference visualization of the anterior angle, called a virtual gonioscopy in a single scan, as shown in Fig.\@ \ref{fig:ac3d}.
The full 3-D tomography also enables quantitative assessment of the angle for all circumferences \cite{Baskaran2013Ophthalmol}.

\subsubsection{Ocular surface investigation}
Among several clinically significant targets at the ocular surface, a filtering bleb structure created by trabeculectomy surgery is considered to be one of the most suitable targets for OCT.
A trabeculectomy is a glaucoma surgery in which the trabecular meshwork is excised in part and a filtering structure for the aqueous humor, the bleb, is created beneath the conjunctiva.
The aqueous humor is then drained into the bleb and this finally reduces the intraocular pressure.

Because the bleb is an artificially created wound in the sclera, it will naturally heal with time.
This undesirable wound healing and the associated scarring results in malfunctioning of the bleb, causing the intraocular pressure to rise again.
Although anti-wound healing drug, mitomycin C, is frequently used to prevent the wound from healing, it does not always successfully stop tissue scarring.
Additional intervention, called bleb revision, including secondary surgery could be necessary if scarring occurs.
To optimize the secondary intervention, we need a proper modality to assess the filtering bleb.

\begin{figure}[tbh]
	\centering\includegraphics{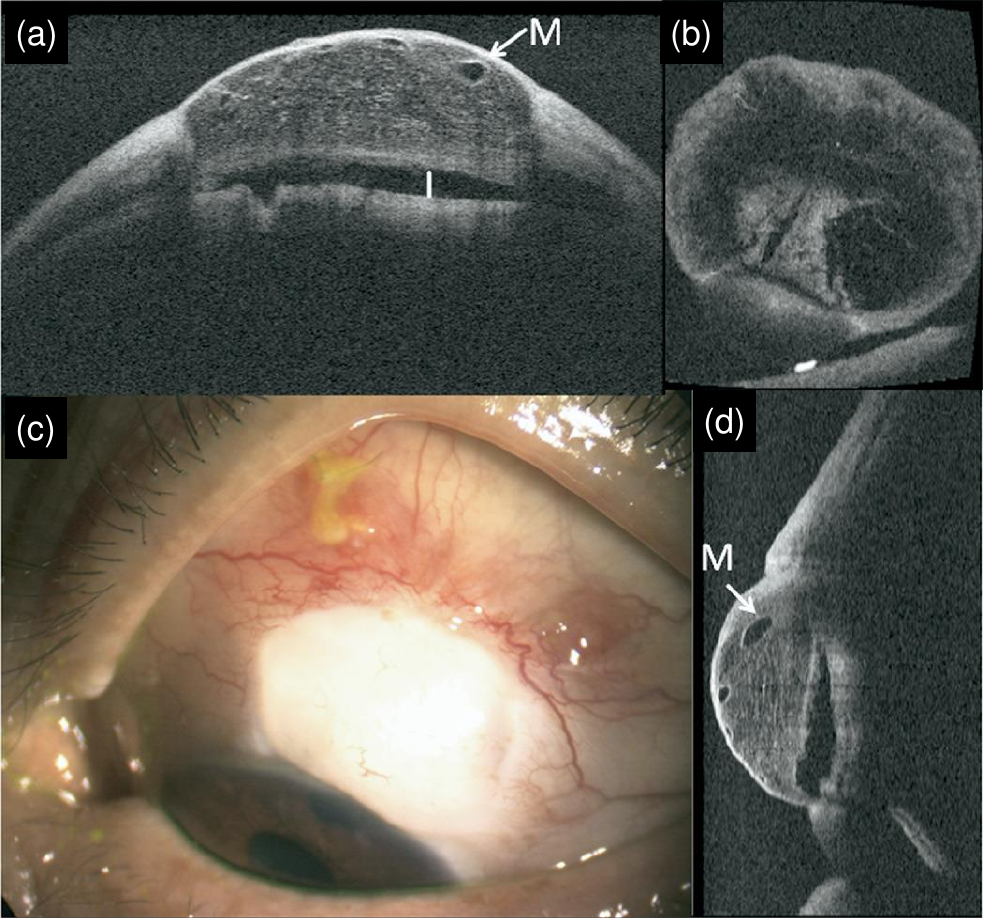}
	\caption{Example of a trabeculectomy bleb visualized by 3-D anterior OCT.
		(c) is a color picture of the bleb, and (a), (b), and (d) are OCT cross-sections in horizontal, vertical, and \enface directions, respectively.
		These OCT cross-sections were extracted from a single volumetric OCT.
		Labels M and I indicate a microcyst and an internal fluid-filled cavity, respectively.}
	\label{fig:bleb}
\end{figure}
Several studies have shown that anterior eye segment OCT at 1310 nm is a suitable modality to assess the filtering bleb \cite{SIngh2007Ophthalmol, Kawana2009Ophthalmol, Kojima2014JG}.
Figure \ref{fig:bleb} (modified and reprinted from Ref.\@ \citen{Kawana2009Ophthalmol}) shows an example of a trabeculectomy bleb examined by 3-D anterior OCT.
Here, the anterior OCT is a 3-D SS-OCT with a 1310-nm probe beam and a depth resolution of 11 \um.
Figures \ref{fig:bleb}(a), (b), and (c) respectively show horizontal, vertical, and \enface OCT cross-sections that were created by a single OCT volume.
A microcyst (M) and fluid-filled cavity (I) are visible in the bleb.
Additionally, the image clearly shows an internal reticulate structure in the bleb that cannot be non-invasively visualized by modalities other than OCT.

\subsubsection{Corneal evaluation by OCT}
Corneal evaluation by OCT has two major purposes: observation of microscopic disorders and quantification of the optical quality, i.e., aberrations in a wide sense.
For the first purpose, both 1310- and 830-nm OCT are used.
Currently, there is no commercial 830-nm OCT device that specializes in anterior investigation.
However, most commercially available posterior 830-nm SD-OCT devices have a measurement mode for the anterior eye.
Because 830-nm OCT, in general, possesses higher resolution than 1310-nm OCT, it provides fine detail of corneal structural disorders.

Some 3-D anterior eye-segment OCT can quantitatively assess the morphological (aberration) properties of the cornea.
This assessment is particularly useful for the diagnosis of keratoconus, a bilateral, progressive, and non-inflammatory disease that is characterized by ectasia and conic deformation of the cornea.
The full 3-D capability of OCT has enabled the automatic morphological analysis of cornea and is known to have a high-sensitivity for keratoconus detection \cite{Fukuda2013BJO}.

\section{Conclusion}
OCT was, from the beginning, intended for clinical investigation.
From a technological perspective, OCT is just a variation of low-coherence interferometry.
If so, it is natural to ask what the main innovation of OCT is.
The answer is in the name: OCT.
In its etymology, interferometry is a measurement method to observe light interference.
Namely, the purpose of interferometry is to measure an interference signal.
Conversely, tomography originates from the Greek word \textit{tomos} that means to slice, and \textit{-graphy} means recording.
As evident in this etymology, the purpose of tomography is to obtain a cross-sectional slice of a subject.
Namely, the low-coherence interferometer was given a clear clinical purpose when it was renamed OCT.

As discussed throughout the manuscript, current OCT is no longer based on a standard low-coherence interferometer (TD-OCT) but on a spectral interferometer.
However, it is still called OCT.
Additionally, the research field of OCT is no longer limited to the field of optics.
It frequently covers photonics, electronics, signal and image processing, and basic and clinical medicine.
More recently, the research field also covers informatics such as machine learning and computer aided diagnosis.
This expansion of OCT can be understood if OCT is not viewed as the name of a technology but the name of an application-oriented research field.

From this point of view, we believe there are still many important research topics in OCT that have not yet been recognized.
Further, there is still room for many researchers to contribute to this field.

\section{Further reading}
This article was written as a minimally short textbook for students and engineers who newly study about OCT, and hence, it covers only important but classic topics.
To further study about the field of OCT, you can consult several good textbooks and articles.

A video lecture series covering nearly the same topic with this tutorial is given by the author, Yasuno, and is available on YouTube \cite{Yasuno2022VideoLecture}.
This video lecture may help you further understand the OCT.

After reading this tutorial, the readers can be ready to consult some seminal papers, such as Ref.\@ \citen{Nassif2004OpEx} for SD-OCT and Ref.\@ \citen{Yun2003OFDI} for SS-OCT.
These early papers are very good summaries of the fundamental points of FD-OCT.
A specific topic of the FD-OCT's sensitivity is intensively discussed in three seminal papers all published in 2003 \cite{Leightgeb2003OpEx, deBore2003OL, Choma2003OpEx}.
You may have a solid understanding about the sensitivity issues by reading all three or two papers.
Even though it might be difficult to understand the topic from a single paper, reading multiple papers helps you comprehend the sensitivity theory.

``Optical coherence tomography -- technology and applications'' edited by Wolfgang Drexler and James G.\@ Fujimoto \cite{OctTechApp2015} is an extensive textbook of OCT which covers almost all important topics of OCT including the basement technology, light source, technical extensions, and also applications.
Especially, the first chapter ``introduction to OCT'' by Fujimoto and Drexler \cite{Fujimoto1995OctBook} and the second chapter ``theory of optical coherence tomography'' by Izatt \etal \cite{Izatt2015OctBook} are recommended as the next steps after reading this tutorial.
These chapters cover almost the same topics with this tutorial, but more details are given.

In order to study individual topics of OCT, you can consult good review papers.
For example, Optical Society of America (currently Optica) organized a feature issue of Biomedical Optics Express ``25 year anniversary of optical coherence tomography'' in 2017 \cite{Izatt2017BOE_OCT25Y}.
This special issue is a collection of good review papers including about
Fourier domain OCT \cite{deBoer2017BOE_FDOCT},
OCT angiography \cite{CLChen2017BOE},
polarization sensitive OCT \cite{deBoer2017BOE_PSOCT},
adaptive-optics OCT \cite{Pircher2017BOE},
OCT-based elastography \cite{Larin2017BOE},
computational method for OCT \cite{YZLiu2017BOE},
\enface OCT microscopy \cite{Thouvenin2017BOE},
high-speed OCT light source \cite{Klein2017BOE},
endoscopic OCT \cite{Gora2017BOE},
intraoperative OCT \cite{CarrascoZevallos2017BOE}, 
intravascular OCT \cite{Bouma2017BOE},
and medical and industrial translation of OCT\cite{Swanson2017BOE}.
In addition to those in this feature issue, there are several review papers of specific topics published in several journals, such as of
OCT angiography \cite{Ang2018GACXO},
polarization sensitive OCT \cite{deBoer2002JBO, Pircher2011PRER, Baumann2017AS}, and
optical coherence elastography \cite{Kennedy2014JSTQE,SWang2015JBP,Kirby2017JBO}.
In addition, the author of this tutorial, Yasuno, recently published a review of multi-contrast Jones-matrix optical coherence tomography \cite{Yasuno2022JMOCT}.

I believe all of the learners are now ready to explore the field of OCT by themselves.
Enjoy studying and enjoy your research!

\bibliographystyle{osajnl}
\bibliography{OctTextbook.bib}

\begin{thebibliography}{10}
\newcommand{\enquote}[1]{``#1''}

\bibitem{Huang1991Science}
D.~Huang, E.~A. Swanson, C.~P. Lin, J.~S. Schuman, W.~G. Stinson, W.~Chang,
  M.~R. Hee, T.~Flotte, K.~Gregory, C.~A. Puliafito, and J.~G. Fujimoto,
  \enquote{Optical coherence tomography,} Science \textbf{254}, 1178 --1181
  (1991).

\bibitem{IntroOCT}
J.~G. Fujimoto and W.~Drexler, \enquote{Introduction to optical coherence
  tomography,} in \enquote{Optical Coherence Tomography Technology and
  Applications,} , J.~G. Fujimoto and W.~Drexler, eds. (Springer, 2008), pp.
  1--45.

\bibitem{Rollins1998OpEx}
A.~Rollins, S.~Yazdanfar, M.~Kulkarni, R.~Ung-Arunyawee, and J.~Izatt,
  \enquote{In vivo video rate optical coherence tomography,} Opt. Express
  \textbf{3}, 219--229 (1998).

\bibitem{Potsaid2010OpEx}
B.~Potsaid, B.~Baumann, D.~Huang, S.~Barry, A.~E. Cable, J.~S. Schuman, J.~S.
  Duker, and J.~G. Fujimoto, \enquote{Ultrahigh speed 1050nm swept source /
  {Fourier} domain {OCT} retinal and anterior segment imaging at 100,000 to
  400,000 axial scans per second,} Opt. Express \textbf{18}, 20029--20048
  (2010).

\bibitem{Klein2013BOE}
T.~Klein, W.~Wieser, L.~Reznicek, A.~Neubauer, A.~Kampik, and R.~Huber,
  \enquote{Multi-{MHz} retinal {OCT},} Biomed. Opt. Express \textbf{4},
  1890--1908 (2013).

\bibitem{Pierce2004JID}
M.~C. Pierce, J.~Strasswimmer, B.~H. Park, B.~Cense, and J.~F. de~Boer,
  \enquote{Advances in {Optical} {Coherence} {Tomography} {Imaging} for
  {Dermatology},} J. Invest. Dermatol. \textbf{123}, 458--463 (2004).

\bibitem{Otis2000JADA}
L.~L. Otis, M.~J. Everett, U.~S. Sathyam, and B.~W. Colston, \enquote{Optical
  {Coherence} {Tomography}: {A} {Nnew} {Imaging} {Technology} for {Dentistry},}
  J. Am. Dent. Assoc. \textbf{131}, 511--514 (2000).

\bibitem{Gora2013NatMed}
{Michalina J. Gora}, J.~S. Sauk, R.~W. Carruth, K.~A. Gallagher, M.~J. Suter,
  N.~S. Nishioka, L.~E. Kava, M.~Rosenberg, B.~E. Bouma, and G.~J. Tearney,
  \enquote{Tethered capsule endomicroscopy enables less invasive imaging of
  gastrointestinal tract microstructure,} Nat. Med. \textbf{19}, 238--240
  (2013).

\bibitem{Vakoc2012NRC}
B.~J. Vakoc, D.~Fukumura, R.~K. Jain, and B.~E. Bouma, \enquote{Cancer imaging
  by optical coherence tomography: preclinical progress and clinical
  potential,} Nat. Rev. Cancer \textbf{12}, 363--368 (2012).

\bibitem{Tearney2008JACCI}
G.~J. Tearney, S.~Waxman, M.~Shishkov, B.~J. Vakoc, M.~J. Suter, M.~I.
  Freilich, A.~E. Desjardins, W.-Y. Oh, L.~A. Bartlett, M.~Rosenberg, and B.~E.
  Bouma, \enquote{Three-{Dimensional} {Coronary} {Artery} {Microscopy} by
  {Intracoronary} {Optical} {Frequency} {Domain} {Imaging},} J. Am. Coll.
  Cardiol. Img. \textbf{1}, 752--761 (2008).

\bibitem{Sakata2009CXO}
L.~M. Sakata, J.~DeLeon-Ortega, V.~Sakata, and C.~A. Girkin, \enquote{Optical
  coherence tomography of the retina and optic nerve – a review,} Clin. Exp.
  Ophthalmol. \textbf{37}, 90--99 (2009).

\bibitem{Swanson1992OL}
E.~A. Swanson, D.~Huang, M.~R. Hee, J.~G. Fujimoto, C.~P. Lin, and C.~A.
  Puliafito, \enquote{High-speed optical coherence domain reflectometry,} Opt.
  Lett. \textbf{17}, 151--153 (1992).

\bibitem{Swanson1993OL}
E.~A. Swanson, J.~A. Izatt, M.~R. Hee, D.~Huang, C.~P. Lin, J.~S. Schuman,
  C.~A. Puliafito, and J.~G. Fujimoto, \enquote{In vivo retinal imaging by
  optical coherence tomography,} Opt. Lett. \textbf{18}, 1864--1866 (1993).

\bibitem{ANSI136.1-2014}
{American National Standard Institute}, \emph{American {National} {Standard}
  for the {Safe} {Use} of {Lasers} {ANSI} Z136.1-2014} (American National
  Standards Institute, New York, 2014).

\bibitem{ISO15004-1-2009}
{CH/172/6}, \emph{Ophthalmic instruments. {Fundamental} requirements and test
  methods. {General} requirements applicable to all ophthalmic instruments,
  {BS} {EN} {ISO} 15004-1:2009} (International Organization for
  Standardization, 2009).

\bibitem{Fercher1995OC}
A.~Fercher, C.~Hitzenberger, G.~Kamp, and S.~El-Zaiat, \enquote{Measurement of
  intraocular distances by backscattering spectral interferometry,} Opt.
  Commun. \textbf{117}, 43--48 (1995).

\bibitem{Hauesler1998JBO}
G.~Ha{\"u}sler, \enquote{``{Coherence} {Radar}'' and ``{Spectral}
  {Radar}''—{New} {Tools} for {Dermatological} {Diagnosis},} J. Biomed. Opt.
  \textbf{3}, 21 (1998).

\bibitem{Yun2003OFDI}
S.~Yun, G.~Tearney, J.~de~Boer, N.~Iftimia, and B.~Bouma, \enquote{High-speed
  optical frequency-domain imaging,} Opt. Express \textbf{11}, 2953--2963
  (2003).

\bibitem{Wojtkowski2002JBO}
M.~Wojtkowski, R.~Leitgeb, A.~Kowalczyk, T.~Bajraszewski, and A.~F. Fercher,
  \enquote{In vivo human retinal imaging by {Fourier} domain optical coherence
  tomography,} J. Biomed. Opt. \textbf{7}, 457 (2002).

\bibitem{Nassif2004OpEx}
N.~Nassif, B.~Cense, B.~Park, M.~Pierce, S.~Yun, B.~Bouma, G.~Tearney, T.~Chen,
  and J.~de~Boer, \enquote{In vivo high-resolution video-rate spectral-domain
  optical coherence tomography of the human retina and optic nerve,} Opt.
  Express \textbf{12}, 367--376 (2004).

\bibitem{Leightgeb2003OpEx}
R.~Leitgeb, C.~Hitzenberger, and A.~Fercher, \enquote{Performance of fourier
  domain vs. time domain optical coherence tomography,} Opt. Express
  \textbf{11}, 889--894 (2003).

\bibitem{deBore2003OL}
J.~F. de~Boer, B.~Cense, B.~H. Park, M.~C. Pierce, G.~J. Tearney, and B.~E.
  Bouma, \enquote{Improved signal-to-noise ratio in spectral-domain compared
  with time-domain optical coherence tomography,} Opt. Lett. \textbf{28},
  2067--2069 (2003).

\bibitem{Choma2003OpEx}
M.~Choma, M.~Sarunic, C.~Yang, and J.~Izatt, \enquote{Sensitivity advantage of
  swept source and {Fourier} domain optical coherence tomography,} Opt. Express
  \textbf{11}, 2183 (2003).

\bibitem{Yasuno2005OpEx}
Y.~Yasuno, V.~D. Madjarova, S.~Makita, M.~Akiba, A.~Morosawa, C.~Chong,
  T.~Sakai, K.-P. Chan, M.~Itoh, and T.~Yatagai, \enquote{Three-dimensional and
  high-speed swept-source optical coherence tomography for in vivo
  investigation of human anterior eye segments,} Opt. Express \textbf{13},
  10652--10664 (2005).

\bibitem{Makita2008OpEx}
S.~Makita, T.~Fabritius, and Y.~Yasuno, \enquote{Full-range, high-speed,
  high-resolution 1-$\mu$m spectral-domain optical coherence tomography using
  {BM}-scan for volumetric imaging of the human posterior eye,} Opt. Express
  \textbf{16}, 8406--8420 (2008).

\bibitem{Dorrer2000JOSAB}
C.~Dorrer, N.~Belabas, J.-P. Likforman, and M.~Joffre, \enquote{Spectral
  resolution and sampling issues in {Fourier}-transform spectral
  interferometry,} J. Opt. Soc. Am. B \textbf{17}, 1795--1802 (2000).

\bibitem{Wojtkowski2004OpEx}
M.~Wojtkowski, V.~Srinivasan, T.~Ko, J.~Fujimoto, A.~Kowalczyk, and J.~Duker,
  \enquote{Ultrahigh-resolution, high-speed, {Fourier} domain optical coherence
  tomography and methods for dispersion compensation,} Opt. Express
  \textbf{12}, 2404--2422 (2004).

\bibitem{Cense2004OpEx}
B.~Cense, N.~Nassif, T.~Chen, M.~Pierce, S.-H. Yun, B.~Park, B.~Bouma,
  G.~Tearney, and J.~de~Boer, \enquote{Ultrahigh-resolution high-speed retinal
  imaging using spectral-domain optical coherence tomography,} Opt. Express
  \textbf{12}, 2435--2447 (2004).

\bibitem{Huber2005OpEx}
R.~Huber, M.~Wojtkowski, K.~Taira, J.~Fujimoto, and K.~Hsu, \enquote{Amplified,
  frequency swept lasers for frequency domain reflectometry and {OCT} imaging:
  design and scaling principles,} Opt. Express \textbf{13}, 3513--3528 (2005).

\bibitem{Oh2005OL}
W.~Y. Oh, S.~H. Yun, G.~J. Tearney, and B.~E. Bouma, \enquote{115 {kHz} tuning
  repetition rate ultrahigh-speed wavelength-swept semiconductor laser,} Opt.
  Lett. \textbf{30}, 3159--3161 (2005).

\bibitem{Yun2003OpEx_SD}
S.~Yun, G.~Tearney, B.~Bouma, B.~Park, and J.~de~Boer, \enquote{High-speed
  spectral-domain optical coherence tomography at 1.3 $\mu$m wavelength,} Opt.
  Express \textbf{11}, 3598--3604 (2003).

\bibitem{EZhang2011OpEx}
E.~Z. Zhang and B.~J. Vakoc, \enquote{Polarimetry noise in fiber-based optical
  coherence tomography instrumentation,} Opt. Express \textbf{19}, 16830--16842
  (2011).

\bibitem{EZhang2013OpEx}
E.~Z. Zhang, W.-Y. Oh, M.~L. Villiger, L.~Chen, B.~E. Bouma, and B.~J. Vakoc,
  \enquote{Numerical compensation of system polarization mode dispersion in
  polarization-sensitive optical coherence tomography,} Opt. Express
  \textbf{21}, 1163--1180 (2013).

\bibitem{Villiger2013OL}
M.~Villiger, E.~Z. Zhang, S.~Nadkarni, W.-Y. Oh, B.~E. Bouma, and B.~J. Vakoc,
  \enquote{Artifacts in polarization-sensitive optical coherence tomography
  caused by polarization mode dispersion,} Opt. Lett. \textbf{38}, 923--925
  (2013).

\bibitem{Braaf2011OpEx}
{Boy Braaf}, K.~A. Vermeer, V.~A.~D. Sicam, E.~van Zeeburg, J.~C. van Meurs,
  and J.~F. de~Boer, \enquote{Phase-stabilized optical frequency domain imaging
  at 1-$\mu$m for the measurement of blood flow in the human choroid,} Opt.
  Express \textbf{19}, 20886--20903 (2011).

\bibitem{Hale1973AO}
G.~M. Hale and M.~R. Querry, \enquote{Optical constants of water in the 200-nm
  to 200-$\mu$m wavelength region,} Appl. Opt. \textbf{12}, 555--563 (1973).

\bibitem{Povazay2003OpEx}
B.~Pova\v{z}ay, K.~Bizheva, B.~Hermann, A.~Unterhuber, H.~Sattmann, A.~Fercher,
  W.~Drexler, C.~Schubert, P.~Ahnelt, M.~Mei, R.~Holzwarth, W.~Wadsworth,
  J.~Knight, and P.~S.~J. Russell, \enquote{Enhanced visualization of choroidal
  vessels using ultrahigh resolution ophthalmic {OCT} at 1050 nm,} Opt. Express
  \textbf{11}, 1980--1986 (2003).

\bibitem{Unterhuber2005OpEx}
A.~Unterhuber, B.~Pova\v{z}ay, B.~Hermann, H.~Sattmann, A.~Chavez-Pirson, and
  W.~Drexler, \enquote{In vivo retinal optical coherence tomography at 1040 nm
  - enhanced penetration into the choroid,} Opt. Express \textbf{13},
  3252--3258 (2005).

\bibitem{Lee2006OpEx}
E.~C. Lee, J.~F. de~Boer, M.~Mujat, H.~Lim, and S.~H. Yun, \enquote{In vivo
  optical frequency domain imaging of human retina and choroid,} Opt. Express
  \textbf{14}, 4403--4411 (2006).

\bibitem{Yasuno2007OpEx}
Y.~Yasuno, Y.~Hong, S.~Makita, M.~Yamanari, M.~Akiba, M.~Miura, and T.~Yatagai,
  \enquote{In vivo high-contrast imaging of deep posterior eye by 1-$\mu$m
  swept source optical coherence tomography andscattering optical coherence
  angiography,} Opt. Express \textbf{15}, 6121--6139 (2007).

\bibitem{Srinivasan2008IOVS}
V.~J. Srinivasan, D.~C. Adler, Y.~Chen, I.~Gorczynska, R.~Huber, J.~S. Duker,
  J.~S. Schuman, and J.~G. Fujimoto, \enquote{Ultrahigh-{Speed} {Optical}
  {Coherence} {Tomography} for {Three}-{Dimensional} and {En} {Face} {Imaging}
  of the {Retina} and {Optic} {Nerve} {Head},} Invest. Ophthalmol. Vis. Sci.
  \textbf{49}, 5103 --5110 (2008).

\bibitem{Povazay2007JBO}
B.~Pova\v{z}ay, B.~Hermann, A.~Unterhuber, B.~Hofer, H.~Sattmann, F.~Zeiler,
  J.~E. Morgan, C.~Falkner-Radler, C.~Glittenberg, S.~Blinder, and W.~Drexler,
  \enquote{Three-dimensional optical coherence tomography at 1050nm versus
  800nm in retinal pathologies: enhanced performance and choroidal penetration
  in cataract patients,} J. Biomed. Opt. \textbf{12}, 041211--041211--7 (2007).

\bibitem{Puvanathasan2008OL}
P.~Puvanathasan, P.~Forbes, Z.~Ren, D.~Malchow, S.~Boyd, and K.~Bizheva,
  \enquote{High-speed, high-resolution {Fourier}-domain optical coherence
  tomography system for retinal imaging in the 1060 nm wavelength region,} Opt.
  Lett. \textbf{33}, 2479--2481 (2008).

\bibitem{Radhakrishnan2001ArchOphthalmol}
S.~Radhakrishnan, A.~M. Rollins, J.~E. Roth, S.~Yazdanfar, V.~Westphal, D.~S.
  Bardenstein, and J.~A. Izatt, \enquote{Real-{Time} {Optical} {Coherence}
  {Tomography} of the {Anterior} {Segment} at 1310 nm,} Arch. Ophthalmol.
  \textbf{119}, 1179--1185 (2001).

\bibitem{OctHandbook}
B.~Bouma, ed., \emph{Handbook of {Optical} {Coherence} {Tomography}} (Informa
  Healthcare, 2001), 1st ed.

\bibitem{Zawadzki2005OpEx}
R.~Zawadzki, S.~Jones, S.~Olivier, M.~Zhao, B.~Bower, J.~Izatt, S.~Choi,
  S.~Laut, and J.~Werner, \enquote{Adaptive-optics optical coherence tomography
  for high-resolution and high-speed 3d retinal in vivo imaging,} Opt. Express
  \textbf{13}, 8532--8546 (2005).

\bibitem{Felberer2014BOE}
F.~Felberer, J.-S. Kroisamer, B.~Baumann, S.~Zotter, U.~Schmidt-Erfurth, C.~K.
  Hitzenberger, and M.~Pircher, \enquote{Adaptive optics {SLO}/{OCT} for 3d
  imaging of human photoreceptors in vivo,} Biomed. Opt. Express \textbf{5},
  439--456 (2014).

\bibitem{Hong2007OpEx}
Y.~Hong, S.~Makita, M.~Yamanari, M.~Miura, S.~Kim, T.~Yatagai, and Y.~Yasuno,
  \enquote{Three-dimensional visualization of choroidal vessels by using
  standard and ultra-high resolution scattering optical coherence angiography,}
  Opt. Express \textbf{15}, 7538--7550 (2007).

\bibitem{Makita2006OpEx}
S.~Makita, Y.~Hong, M.~Yamanari, T.~Yatagai, and Y.~Yasuno, \enquote{Optical
  coherence angiography,} Opt. Express \textbf{14}, 7821--7840 (2006).

\bibitem{Tan2008Ophthalmol}
O.~Tan, G.~Li, A.~T.-H. Lu, R.~Varma, and D.~Huang, \enquote{Mapping of
  {Macular} {Substructures} with {Optical} {Coherence} {Tomography} for
  {Glaucoma} {Diagnosis},} Ophthalmology \textbf{115}, 949--956 (2008).

\bibitem{Yasuno2009IOVS}
Y.~Yasuno, M.~Miura, K.~Kawana, S.~Makita, M.~Sato, F.~Okamoto, M.~Yamanari,
  T.~Iwasaki, T.~Yatagai, and T.~Oshika, \enquote{Visualization of
  {Sub}-retinal {Pigment} {Epithelium} {Morphologies} of {Exudative} {Macular}
  {Diseases} by {High}-{Penetration} {Optical} {Coherence} {Tomography},}
  Invest. Ophthalmol. Vis. Sci. \textbf{50}, 405 --413 (2009).

\bibitem{Chung2011Ophthalmol}
S.~E. Chung, S.~W. Kang, J.~H. Lee, and Y.~T. Kim, \enquote{Choroidal
  {Thickness} in {Polypoidal} {Choroidal} {Vasculopathy} and {Exudative}
  {Age}-related {Macular} {Degeneration},} Ophthalmology \textbf{118}, 840--845
  (2011).

\bibitem{Maruko2011Retina}
I.~Maruko, T.~Iida, Y.~Sugano, A.~Ojima, and T.~Sekiryu, \enquote{Subfoveal
  choroidal thickness in fellow eyes of patients with central serous
  chorioretinopathy,} Retina \textbf{31}, 1603--1608 (2011).

\bibitem{Kuroda2013Retina}
S.~Kuroda, Y.~Ikuno, Y.~Yasuno, K.~Nakai, S.~Usui, M.~Sawa, M.~Tsujikawa,
  F.~Gomi, and K.~Nishida, \enquote{Choroidal thickness in central serous
  chorioretinopathy,} Retina \textbf{33}, 302--308 (2013).

\bibitem{Usui2012AJO}
S.~Usui, Y.~Ikuno, A.~Miki, K.~Matsushita, Y.~Yasuno, and K.~Nishida,
  \enquote{Evaluation of the {Choroidal} {Thickness} {Using}
  {High}-{Penetration} {Optical} {Coherence} {Tomography} {With} {Long}
  {Wavelength} in {Highly} {Myopic} {Normal}-{Tension} {Glaucoma},} Am. J.
  Ophthalmol. \textbf{153}, 10--16.e1 (2012).

\bibitem{Fujiwara2009AJO}
T.~Fujiwara, Y.~Imamura, R.~Margolis, J.~S. Slakter, and R.~F. Spaide,
  \enquote{Enhanced {Depth} {Imaging} {Optical} {Coherence} {Tomography} of the
  {Choroid} in {Highly} {Myopic} {Eyes},} Am. J. Ophthalmol. \textbf{148},
  445--450 (2009).

\bibitem{Ikuno2013CO}
Y.~Ikuno, S.~Fujimoto, Y.~Jo, T.~Asai, and K.~Nishida, \enquote{Choroidal
  thinning in high myopia measured by optical coherence tomography,} Clin.
  Ophthalmol. \textbf{7}, 889--893 (2013).

\bibitem{Pavlin1992AJO}
C.~J. Pavlin, K.~Harasiewicz, and F.~S. Foster, \enquote{Ultrasound
  biomicroscopy of anterior segment structures in normal and glaucomatous
  eyes.} Am. J. Ophthalmol. \textbf{113}, 381--389 (1992).

\bibitem{Ishikawa2000COO}
H.~Ishikawa, J.~M. Liebmann, and R.~Ritch, \enquote{Quantitative assessment of
  the anterior segment using ultrasound biomicroscopy.} Curr. Opin. Ophthalmol.
  \textbf{11}, 133--139 (2000).

\bibitem{Baskaran2013Ophthalmol}
M.~Baskaran, S.-W. Ho, T.~A. Tun, A.~C. How, S.~A. Perera, D.~S. Friedman, and
  T.~Aung, \enquote{Assessment of {Circumferential} {Angle}-{Closure} by the
  {Iris}–{Trabecular} {Contact} {Index} with {Swept}-{Source} {Optical}
  {Coherence} {Tomography},} Ophthalmology \textbf{120}, 2226--2231 (2013).

\bibitem{SIngh2007Ophthalmol}
M.~Singh, P.~T.~K. Chew, D.~S. Friedman, W.~P. Nolan, J.~L. See, S.~D. Smith,
  C.~Zheng, P.~J. Foster, and T.~Aung, \enquote{Imaging of {Trabeculectomy}
  {Blebs} {Using} {Anterior} {Segment} {Optical} {Coherence} {Tomography},}
  Ophthalmology \textbf{114}, 47--53 (2007).

\bibitem{Kawana2009Ophthalmol}
K.~Kawana, T.~Kiuchi, Y.~Yasuno, and T.~Oshika, \enquote{Evaluation of
  {Trabeculectomy} {Blebs} {Using} 3-{Dimensional} {Cornea} and {Anterior}
  {Segment} {Optical} {Coherence} {Tomography},} Ophthalmology \textbf{116},
  848--855 (2009).

\bibitem{Kojima2014JG}
S.~Kojima, T.~Inoue, T.~Kawaji, and H.~Tanihara, \enquote{Filtration {Bleb}
  {Revision} {Guided} by 3-{Dimensional} {Anterior} {Segment} {Optical}
  {Coherence} {Tomography},} J. Glaucoma \textbf{23}, 312--315 (2014).

\bibitem{Fukuda2013BJO}
S.~Fukuda, S.~Beheregaray, S.~Hoshi, M.~Yamanari, Y.~Lim, T.~Hiraoka,
  Y.~Yasuno, and T.~Oshika, \enquote{Comparison of three-dimensional optical
  coherence tomography and combining a rotating {Scheimpflug} camera with a
  {Placido} topography system for forme fruste keratoconus diagnosis,} Br. J.
  Ophthalmol. \textbf{97}, 1554--1559 (2013).

\bibitem{Yasuno2022VideoLecture}
Y.~Yasuno, \enquote{Fundamentals of {Fourier}-domain optical coherence
  tomography,} Video lecture series on YouTube (2022).
  {\url{https://youtube.com/playlist?list=PL38KibqB_aSATnTfh_ehcsSG00IUMIlQq}}.

\bibitem{OctTechApp2015}
W.~Drexler and J.~G. Fujimoto, eds., \emph{Optical Coherence Tomography ---
  technology and applications} (Springer International Publishing, 2015), 2nd
  ed.

\bibitem{Fujimoto1995OctBook}
J.~G. Fujimoto and W.~Drexler, \emph{Introduction to OCT} (Springer
  International Publishing, Cham, 2015), pp. 3--64.

\bibitem{Izatt2015OctBook}
J.~A. Izatt, M.~A. Choma, and A.-H. Dhalla, \emph{Theory of Optical Coherence
  Tomography} (Springer International Publishing, Cham, 2015), pp. 65--94.

\bibitem{Izatt2017BOE_OCT25Y}
J.~A. Izatt, S.~Boppart, B.~Bouma, J.~de~Boer, W.~Drexler, X.~Li, and
  Y.~Yasuno, \enquote{Introduction to the feature issue on the 25 year
  anniversary of optical coherence tomography,} Biomed. Opt. Express
  \textbf{8}, 3289--3291 (2017).

\bibitem{deBoer2017BOE_FDOCT}
J.~F. de~Boer, R.~Leitgeb, and M.~Wojtkowski, \enquote{Twenty-five years of
  optical coherence tomography: the paradigm shift in sensitivity and speed
  provided by fourier domain oct [invited],} Biomed. Opt. Express \textbf{8},
  3248--3280 (2017).

\bibitem{CLChen2017BOE}
C.-L. Chen and R.~K. Wang, \enquote{Optical coherence tomography based
  angiography [invited],} Biomed. Opt. Express \textbf{8}, 1056--1082 (2017).

\bibitem{deBoer2017BOE_PSOCT}
J.~F. de~Boer, C.~K. Hitzenberger, and Y.~Yasuno, \enquote{Polarization
  sensitive optical coherence tomography --- a review [invited],} Biomed. Opt.
  Express \textbf{8}, 1838--1873 (2017).

\bibitem{Pircher2017BOE}
M.~Pircher and R.~J. Zawadzki, \enquote{Review of adaptive optics oct (ao-oct):
  principles and applications for retinal imaging [invited],} Biomed. Opt.
  Express \textbf{8}, 2536--2562 (2017).

\bibitem{Larin2017BOE}
K.~V. Larin and D.~D. Sampson, \enquote{Optical coherence elastography --- oct
  at work in tissue biomechanics [invited],} Biomed. Opt. Express \textbf{8},
  1172--1202 (2017).

\bibitem{YZLiu2017BOE}
Y.-Z. Liu, F.~A. South, Y.~Xu, P.~S. Carney, and S.~A. Boppart,
  \enquote{Computational optical coherence tomography [invited],} Biomed. Opt.
  Express \textbf{8}, 1549--1574 (2017).

\bibitem{Thouvenin2017BOE}
O.~Thouvenin, K.~Grieve, P.~Xiao, C.~Apelian, and A.~C. Boccara,
  \enquote{\textit{En face} coherence microscopy [invited],} Biomed Opt.
  Express. \textbf{8}, 622--639 (2017).

\bibitem{Klein2017BOE}
T.~Klein and R.~Huber, \enquote{High-speed oct light sources and systems
  [invited],} Biomed. Opt. Express \textbf{8}, 828--859 (2017).

\bibitem{Gora2017BOE}
M.~J. Gora, M.~J. Suter, G.~J. Tearney, and X.~Li, \enquote{Endoscopic optical
  coherence tomography: technologies and clinical applications [invited],}
  Biomed. Opt. Express \textbf{8}, 2405--2444 (2017).

\bibitem{CarrascoZevallos2017BOE}
O.~M. Carrasco-Zevallos, C.~Viehland, B.~Keller, M.~Draelos, A.~N. Kuo, C.~A.
  Toth, and J.~A. Izatt, \enquote{Review of intraoperative optical coherence
  tomography: technology and applications [invited],} Biomed. Opt. Express
  \textbf{8}, 1607--1637 (2017).

\bibitem{Bouma2017BOE}
B.~E. Bouma, M.~Villiger, K.~Otsuka, and W.-Y. Oh, \enquote{Intravascular
  optical coherence tomography [invited],} Biomed. Opt. Express \textbf{8},
  2660--2686 (2017).

\bibitem{Swanson2017BOE}
E.~A. Swanson and J.~G. Fujimoto, \enquote{The ecosystem that powered the
  translation of oct from fundamental research to clinical and commercial
  impact [invited],} Biomed. Opt. Express \textbf{8}, 1638--1664 (2017).

\bibitem{Ang2018GACXO}
M.~Ang, A.~C.~S. Tan, C.~M.~G. Cheung, P.~A. Keane, R.~Dolz-Marco, C.~C.~A.
  Sng, and L.~Schmetterer, \enquote{Optical coherence tomography angiography: a
  review of current and future clinical applications,} Graefes Arch. Clin. Exp.
  Ophthalmol. \textbf{256}, 237--245 (2018).

\bibitem{deBoer2002JBO}
J.~F. de~Boer and T.~E. Milner, \enquote{Review of polarization sensitive
  optical coherence tomography and stokes vector determination,} J. Biomed.
  Opt. \textbf{7}, 359 -- 371 (2002).

\bibitem{Pircher2011PRER}
M.~Pircher, C.~K. Hitzenberger, and U.~Schmidt-Erfurth, \enquote{Polarization
  sensitive optical coherence tomography in the human eye,} Prog. Retin. Eye
  Res. \textbf{30}, 431--451 (2011).

\bibitem{Baumann2017AS}
B.~Baumann, \enquote{Polarization sensitive optical coherence tomography: A
  review of technology and applications,} Appl. Sci. \textbf{7} (2017).

\bibitem{Kennedy2014JSTQE}
B.~F. Kennedy, K.~M. Kennedy, and D.~D. Sampson, \enquote{A review of optical
  coherence elastography: Fundamentals, techniques and prospects,} IEEE J. Sel.
  Top. Quantum Electron. \textbf{20}, 272--288 (2014).

\bibitem{SWang2015JBP}
S.~Wang and K.~V. Larin, \enquote{Optical coherence elastography for tissue
  characterization: a review,} J Biophotonics \textbf{8}, 279--302 (2015).

\bibitem{Kirby2017JBO}
M.~A. Kirby, I.~Pelivanov, S.~Song, L.~Ambrozinski, S.~J. Yoon, L.~Gao, D.~Li,
  T.~T. Shen, R.~K. Wang, and M.~O'Donnell, \enquote{{Optical coherence
  elastography in ophthalmology},} J. Biomed. Opt. \textbf{22}, 121720 (2017).

\bibitem{Yasuno2022JMOCT}
Y.~Yasuno, \enquote{Multi-contrast {Jones}-matrix optical coherence tomography
  - the concept, principle, implementation, and applications,}  (2022).
  ArXiv:2211.17151 [physics].

\end{thebibliography}

\section*{Change Log}
\begin{itemize}
	\item 2014-07-14 Version 2.0a: The first online version released.
	\item 2022-12-05 Version 3.0pre01: Reformatted for a creative commons version.
    \item 2022-12-06 Version 3.0pre02: Figures were colorized and captions were extended.
    \item 2022-12-08 Version 3.0a: A new section ``Further reading'' was added.
\end{itemize}

 \end{document}